\documentclass[smallextended]{svjour3}       
\smartqed  
\usepackage{graphicx}
\usepackage[english]{babel}

 \usepackage{amsmath}
 \usepackage{amsfonts}
 \usepackage[T1]{fontenc}

\usepackage{amssymb}
\usepackage{graphicx}
\usepackage{dsfont} 

\usepackage{color}
\usepackage{array}
\usepackage{supertabular}
\usepackage{hhline}
\usepackage{multirow}
\usepackage[table]{xcolor}

\usepackage{float}

\usepackage{longtable}

\usepackage{natbib}

\usepackage{stfloats}

\usepackage{fancyhdr}
\definecolor{rltred}{rgb}{0.75,0,0}
\definecolor{rltgreen}{rgb}{0,0.5,0}
\definecolor{forestgreen}{rgb}{0.13,0.54,0.13}
\definecolor{rltblue}{rgb}{0,0,0.75}
\definecolor{rltblack}{rgb}{0,0,0}

\usepackage[
   extension=pdf,           
   colorlinks,
   urlcolor=rltblue,       
   citecolor=rltblue ,  
   filecolor=rltblue,     
   linkcolor=rltblue,       
   pdftitle={Titre du document},
   pdfauthor={CA},
   pdfsubject={},
   pdfkeywords={},
   pdfproducer={pdflatex},%
   pdfpagemode=None,
   bookmarksopen=1,
] {hyperref}

\newcommand{\refeq}[1]{(\ref{#1})}

\newenvironment{pfof}[1]{\vspace{1ex}\noindent{\bf Proof of #1:}\hspace{0.5em}}
	{\hfill\qed\vspace{1ex}}

\journalname{}
\begin{document}

\title{A functional analysis of speed profiles: smoothing using derivative information, curve registration, and functional boxplot}



\author{C. Andrieu         \and
        G. Saint Pierre \and
        X. Bressaud
}


\institute{C. Andrieu \at
           French Institute of Science and Technology for Transport, Development and Networks, Laboratory for Vehicle Infrastructure Driver Interactions (IFSTTAR/LIVIC), 77, rue des Chantiers, 78000 Versailles, France \\
              \email{cindie.andrieu@ifsttar.fr}           
           \and
           G. Saint Pierre \at
           French Institute of Science and Technology for Transport, Development and Networks, Laboratory for Vehicle Infrastructure Driver Interactions (IFSTTAR/LIVIC), 77, rue des Chantiers, 78000 Versailles, France \\
           \email{guillaume.saintpierre@ifsttar.fr}
           \and
           X. Bressaud \at
           Universit\'{e} Paul Sabatier, Institut de Math\'{e}matiques de Toulouse, F-31062 Toulouse Cedex 9, France \\
           \email{bressaud@math.univ-toulouse.fr}
}

\date{}

\maketitle

\begin{abstract}
In this paper, we propose a functional analysis of a set of individual space-speed profiles corresponding to speed as function of the distance traveled by the vehicle from an initial point. This functional analysis begins with a functional modeling of space-speed profiles and the study of mathematical properties of these functions. Then, in a first step, a smoothing procedure based on spline smoothing is developed in order to convert the raw data into functional objets and to filter out the measurement noise as efficiently as possible. It is shown that this smoothing step leads to a complex nonparametric regression problem that needs to take into account two constraints: the use of the derivative information, and a monotonicity constraint. The performance of the proposed two-step estimator (smooth, and then monotonize) is illustrated on simulation studies and a real data example. In a second step, we use a curve registration method based on landmarks alignment in order to construct an average speed profile representative of a set of individual speed profiles. Finally, the variability of such a set is explored by the use of functional boxplots.

\end{abstract}

\section{Introduction}
\label{sect_intro}

The knowledge of the actual vehicle speeds on roads is essential from several points of view: to locate blackspot in the network, to improve the knowledge of travel time and to evaluate the effects of the modification of the infrastructure (addition of speed bumps, roundabouts, ...). The speed choice of drivers is one of the most important components of their behavior and also their road usage. This continuous information of road user's speed is available with the development of probe vehicles, that can be seen as mobile sensors exploring continuously the road network. More particularly, the development of smartphones equipped with a GPS (Global Positioning System) has increased the number of digital "traces" left by vehicles, and leads to the obtention of individual space-speed profiles that represent speed as a function of vehicle position. \par
The collection of individual space-speed profiles can leads to large volume of data that require the use of appropriate methods. Indeed, since in practice space-speed profiles are composed of time-stamped measurements of speed and position, most studies consider them as $\mathds{R}^{n}$ vectors where $n$ is the number of measurements. However, advances in sensors technology enable to collect data with high sampling rate that leads to high dimensional vectors ($n$ is very large), for which classical multivariate statistical methods become inadequate because of problems related to the so-called "curse of dimensionality" and the significant correlation between close observations. \par
The originality of the approach presented in the current paper is to propose a functional analysis of space-speed profiles, i.e. to treat these objects as functions rather than vectors. This approach takes inspiration from Functional Data Analysis, a statistical domain that has developed considerably over the last twenty years and that appears in several domains such as meteorology, chemometrics or economics (e.g. \citealt{Ramsay2002}; \citealt{Febrero2007}), but that is not yet widely used in road transport. Yet the functional approach is particularly suitable for the analysis of speed profiles since it allows to preserve the physical consistency between speed and position (and implicitly time), and their functional characteristics: computation of derivatives (that leads to acceleration or jerk profiles), regularity, shape constraints... An overview of the theory of statistics with functional data can be found in the monographs by \cite{Ramsay2002,Ramsay2005} or \cite{Ferraty2006}, and in the reviews by \cite{Levitin2007}, \cite{Valderrama2007} or more recently \cite{Cuevas2013}. Thus, after a description of a real data set composed of 78 individual space-speed profiles, we begin our analysis by a functional modeling of space-speed profiles with a definition of the corresponding functional space and the study of some mathematical properties (continuity and differentiability). \par
Then, the first step in the functional analysis of space-speed profiles is to convert the raw data including speed and position measurements into functional objects. It is shown that this smoothing problem can be viewed as an interesting nonparametric regression problem that needs to take into account two constraints: the use of the derivative information, and a monotonicity constraint. In this paper, we propose a two-step estimator : smoothing step, and then monotonization step. We show that the smoothing step with the constraint of the use of the derivative information can be seen as a special case of the general spline smoothing problem (see \citealt{Wahba1990}; \citealt{Wang2011}) and can be solved by using the theory of reproducing kernel Hilbert spaces. Thus, the estimator proposed in this paper can be written as a linear combination of basis functions and kernel functions. However, from a computational point of view, we show that the use of a semi-kernel in place of reproducing kernel is more appropriate, and we propose to use the theory of thin-plate spline (\citealt{Wahba1990}; \citealt{Wahba1980}) in order to obtain an estimator with a simpler form. Then, a monotonization step is proposed based on a method developed by \cite{Ramsay1998} which has the advantage of being relatively simple to implement.\par
However, if this smoothing procedure leads to a set of individual space-speed profiles, when the volume of data is large, it is necessary to summarize the information contained in this set. So, in a second time, we proposed a methodology of construction of an aggregated speed profile, such as the average profile. It is then necessary to use curve registration method in order to correct phase variation (especially at stops' location), and to obtain a representative speed profile with similar features of corresponding individual speed profiles. \par
Finally, in a third time, we propose to apply the functional boxplot developed by \cite{Sun2011} using an appropriate functional depth to the set of individual space-speed profiles. This graphical tool which is an extension of the classical boxplot used in the univariate setting, is very interesting to explore the variability of a functional data set. The application of this tool to speed profiles data set leads to the construction of speed corridors that reflect the variability between road users and are very informative about actual operating speed. This speed corridors are particularly adapted to driving assistance system and to enrich or update digital maps.\par
The remainder of the paper is structured as follows. In Sect.~\ref{sect_the_data}, the real data set of individual space-speed profiles used in this study is presented. In Sect.~\ref{sect_functional_modeling_of_ssp}, we propose a definition of the functional space of space-speed profiles and we study some mathematical properties (continuity and differentiability) of these functions. In Sect.~\ref{sect_smoothing_problem}, we propose a two-step smoothing procedure : smooth using derivative information, and then monotonize. Performance and limitations of the estimator are discussed on simulation studies and the real data set. In Sect.~\ref{sect_registration}, we propose a methodology of construction of an average speed profile using a curve registration procedure based on the method of landmarks alignment. In Sect.~\ref{sect_functional_boxplot}, we apply the functional boxplot to the real data set of individual speed-profiles and show the interest of these speed corridors to explore the variability between road users. Finally, Sect.~\ref{sect_conclusion} provides the main conclusions of the present study.

\section{The data}
\label{sect_the_data}

\begin{figure}[!h]
\begin{center}
\includegraphics[width=10cm]{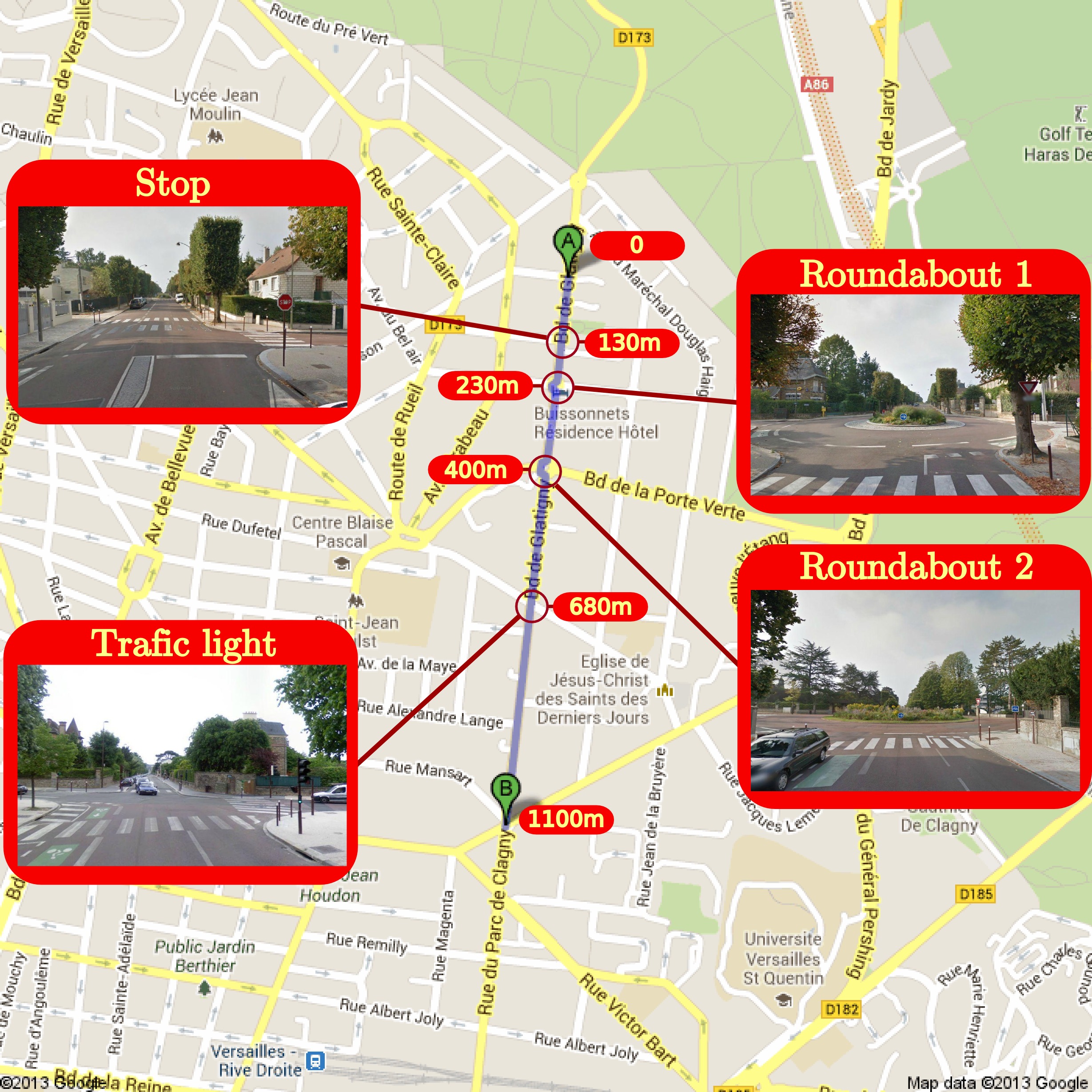}
\end{center}
\caption{Map of the studied section.}
\label{fig_details_section_etudiee_these}
\end{figure}

We consider in this paper a data set extracted from an experiment conducted by the French laboratory IFSTTAR-LIVIC and that took place in 2012 in Versailles, France. Thirty-nine drivers participated to this experiment and performed twice a road section of urban and inter-urban type with a length of about 1100~m. This road section, illustrated at Fig. \ref{fig_details_section_etudiee_these}, corresponds to the path from A to B and is composed of a stop sign, two roundabouts and a traffic light. For logistical reasons, two vehicles were used for this experiment: a Renault Clio III equipped with a Garmin GPS 16x LVC (for 20 drivers), and a Renault Modus with a GPS GlobalSat BR-355 (for 19 drivers). Note that the use of two vehicles and the fact that each driver performed twice the studied section lead to conditions close to naturalistic driving studies where different drivers were observed in a natural setting, in particular during regular travels such as the commute to work. Thus, we do not take into account the correlation between the two paths of the same driver.

~\\

The data-logger collects vehicle position (latitude and longitude) and speed via GPS at a 1Hz sampling frequency (i.e. 1mes/sec). Note that GPS receivers use the Doppler shifts of the satellite signals to calculate vehicle speed, which implies that vehicle speed is independent of vehicle position. In order to reduce the GPS position measurements to a one-dimensional framework, it is assumed that these are map-matched, so that the vehicle is positioned on the correct road segment. Many map-matching algorithms have been developed to identify the correct road segment on which the vehicle is travelling. Thus, in this paper, GPS measurements represent the curvilinear abscissa of the vehicle on the studied road segment (absolute location) from the initial position (point A).\par
\begin{figure}[!h]
\begin{center}
\includegraphics[width=12cm]{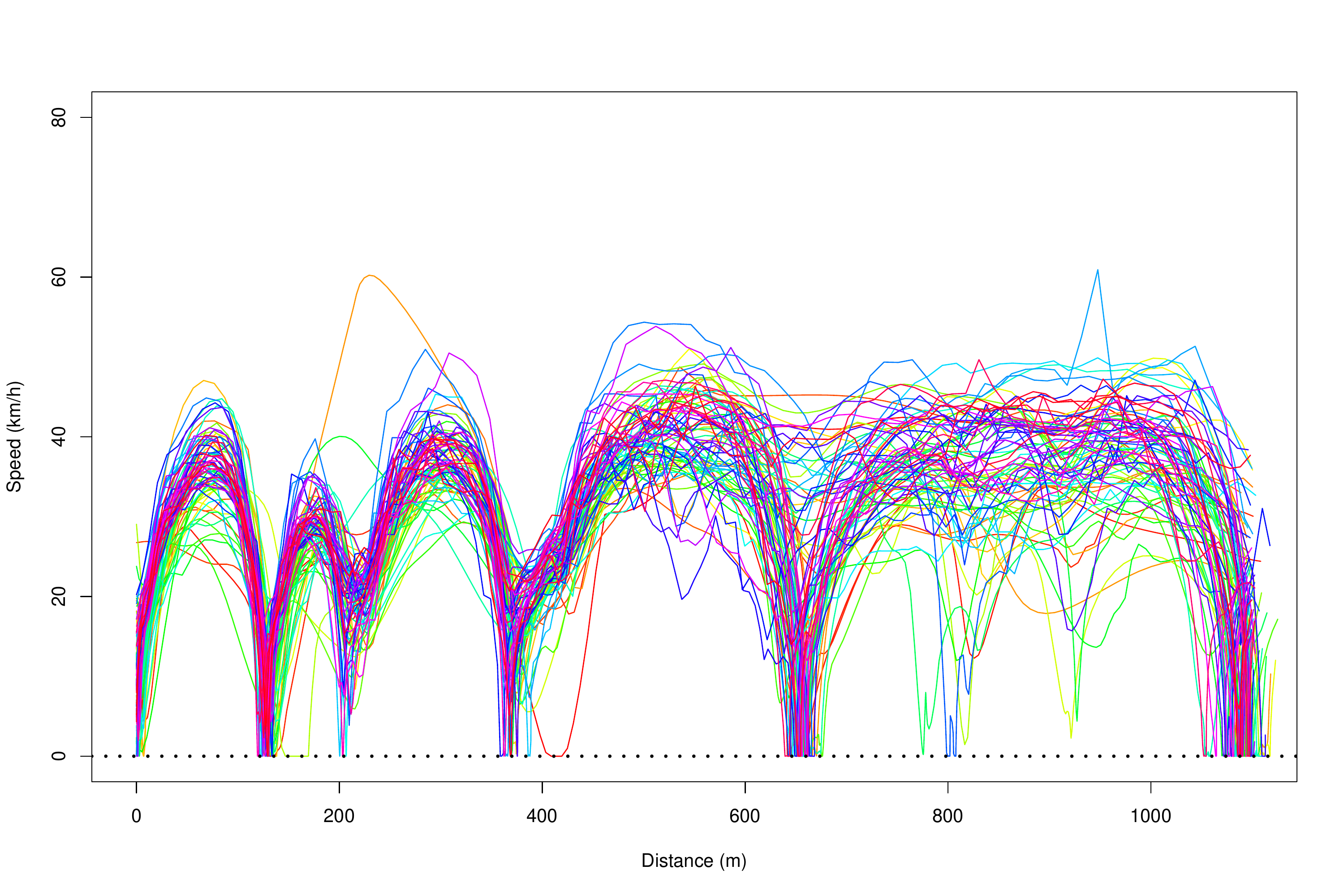} \\
\end{center}
\caption{Sample of 78 space-speed profiles : raw data.}
\label{fig_speed_vs_distance_raw_data}
\end{figure}

The aim of the study is to focus on space-speed profiles from this data set, i.e. speed as a function of the distance traveled by the vehicle from the point A. Since the 39 drivers performed twice the studied section, the data set is composed of 78 individual space-speed profiles illustrated at Fig. \ref{fig_speed_vs_distance_raw_data}.

~\\

\section{Functional modeling of space-speed profiles}
\label{sect_functional_modeling_of_ssp}

\subsection{Definition of space-speed profiles}
\label{subsect_definition_of_ssp}

\begin{figure}[htbp]
\centering
\includegraphics[width=16cm]{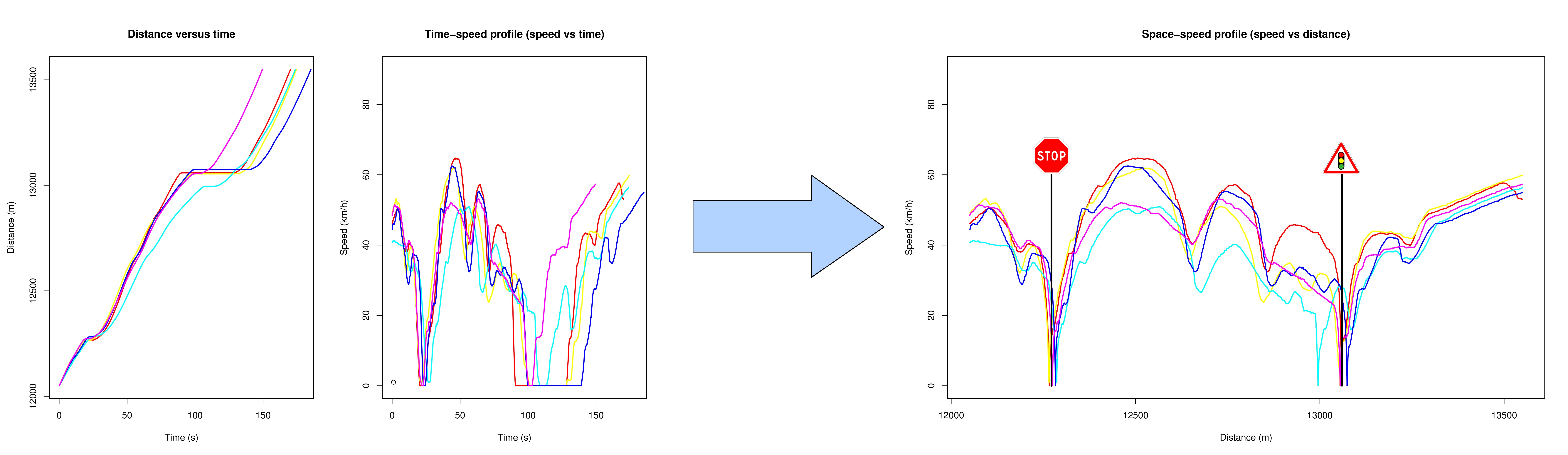}
\caption{Link between the three study areas : [\texttt{distance $\times$ time}, \texttt{speed $\times$ time}] and [\texttt{speed $\times$ distance}].}
\label{fig_lien_entre_espaces}
\end{figure}

Before beginning a functional analysis of space-speed profiles, it is necessary to define the functional space of such objects. Indeed, any function $f : \mathds{R}^{+} \longrightarrow  \mathds{R}^{+}$ is not a space-speed profile (e.g. a constant function equal to zero). In practice, a space-speed profile is a sequence of time-stamped measurements of position (from GPS or odometer) and speed, so it can be studied in the three following study areas: \texttt{distance $\times$ time}, \texttt{speed $\times$ time} and \texttt{speed $\times$ distance} (see Fig. \ref{fig_lien_entre_espaces}).

The functions defined in each of these three study areas are related mathematically: if we denote $F(t)$ a function defined in the study area \texttt{distance $\times$ time} that represents the distance traveled as function of time, the derivative function $F'(t)$ represents the speed as function of time and is defined in the study area \texttt{speed $\times$ time}. So, by definition, the function $F$ must be increasing and at least of class $\mathcal{C}^{1}$. In order to define an acceleration profile, we propose to require that $F$ is at least $\mathcal{C}^{2}$, and so we propose the following definition of the functional space of space-speed profiles :

\begin{definition}
\label{def_space_speed_profiles}
Let $x_{f}\in\mathds{R}^{+}$. Then the space of space-speed profiles, denoted $\mathcal{E}_{SSP}$, is defined as follows :\\
$\mathcal{E}_{SSP}=\{v_{S}:[0,x_{f}]\longrightarrow\mathds{R}^{+}$ such that there exists a positive real T and an increasing function $F:[0,T]\longrightarrow [0,x_{f}]$ of class $\mathcal{C}^{2}$ with $F(0)=0$ such that $v_{S}(x)=F' \circ F^{-1}(x))$, $x\in [0,x_{f}]\}$, \\
where $F^{-1}$ is the generalized inverse of $F$ defined by $F^{-1}(x)=inf\{t\in [0,T],F(t)=x\}$.
\end{definition}

\begin{figure}[htbp]
\centering
\includegraphics[width=2in]{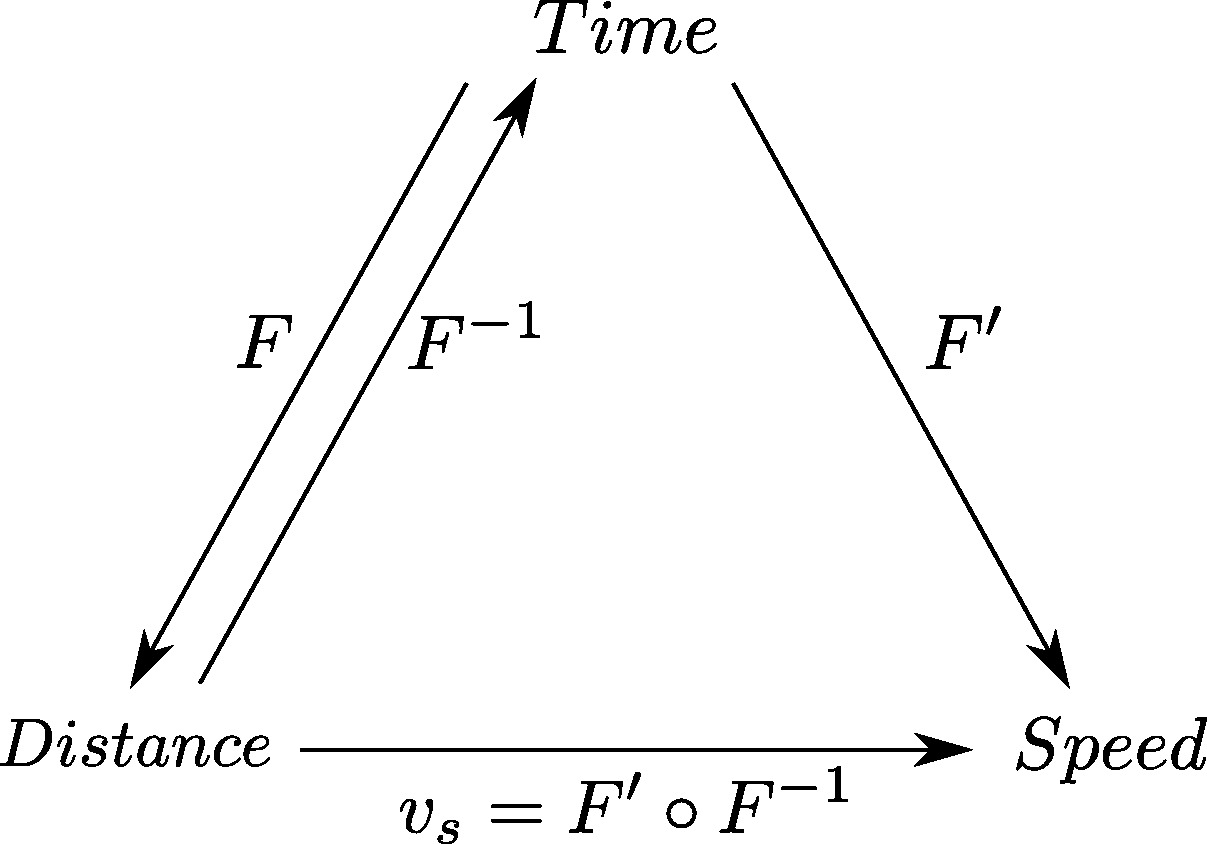}
\caption{Functional diagram illustrating the definition of space-speed profiles.}
\label{fig_diagramme_fonctionnel}
\end{figure}

The positive real numbers $x_{f}$ and $T$ represent respectively the length and the travel time of the studied section. Fig. \ref{fig_diagramme_fonctionnel} illustrates Definition \ref{def_space_speed_profiles} by showing the functional link between distance, speed and implicitly time.

\subsection{Mathematical properties of space-speed profiles}
\label{subsect_properties_of_ssp}

We studied some properties of the space-speed profiles, i.e. functions in the space $\mathcal{E}_{SSP}$ (as defined in Definition \ref{def_space_speed_profiles}). The continuity property is given by the following theorem whose proof is deferred to Appendix \ref{appendix_proofs_properties_of_ssp}:

\begin{theorem}
\label{th_continuity_ssp}
All functions $v_{S}:[0,x_{f}]\longrightarrow\mathds{R}^{+}$ belonging to the space of space-speed profiles $\mathcal{E}_{SSP}$ (as defined in Definition \ref{def_space_speed_profiles}) are continuous on $[0,x_{f}]$.
\end{theorem}

If the continuity property of space-speed profiles is obvious, the differentiability property is less intuitive as shown in the following theorem whose proof is also deferred to Appendix \ref{appendix_proofs_properties_of_ssp}:

\begin{theorem}
\label{th_differentiability_ssp}
Assume that $v_{S}:[0,x_{f}]\longrightarrow\mathds{R}^{+}$ belongs to the space of space-speed profiles $\mathcal{E}_{SSP}$ (as defined in Definition \ref{def_space_speed_profiles}). Let $H_{0}=\{x \in [0,x_{f}], v_{S}(x)=0\}$, all points for which the speed is zero. The two following assumptions are added:
\begin{enumerate}
  \item[$(H_{1})$] Assume that $F$ is of class $\mathcal{C}^{2}$ on $[0,T]$ and strictly increasing, and $\exists t_{0} \in ]0,T[$ such that $F'(t_{0})~=~0$ and $F'''(t_{0})$ exists with $F'''(t_{0}) \neq 0$.
  \item[$(H_{2})$] Assume that $F$ is of class $\mathcal{C}^{2}$ on $[0,T]$ and increasing, $\exists t_{0},t_{1} \in ]0,T[$, $t_{0}\neq t_{1}$ such that $F'(t)=0$ on $[t_{0},t_{1}]$, and the function $G$ defined on $[0,T-(t_{1}-t_{0})]$ by:
      \begin{equation*}
      \left\{
      \begin{array}{l}
       \text{for}\ t\leq t_{0},\ \ G(t)=F(t), \\
       \text{for}\ t\geq t_{0},\ \ G(t)=F(t+t_{1}-t_{0}),
     \end{array}
     \right.
     \end{equation*}
     satisfies the assumptions $(H_{1})$.
\end{enumerate}
If $F$ satisfies the assumptions $(H_{1})$ or $(H_{2})$, then $v_{S}=F' \circ F^{-1}$ is not differentiable on $H_{0}$.
\end{theorem}

The assumptions $(H_{1})$ and $(H_{2})$ are not restrictive and are satisfied in most cases. So, this theorem shows that space-speed profiles are not differentiable at points for which the speed is zero, i.e. when the vehicle is stopped. From a geometrical point of view, if we assume that $x_{0} \in H_{0}$ (i.e. $v_{S}(x_{0})=0$), it is easily shown that the graph of a space-speed profile $v_{S}$ has a half-tangent parallel to the axis of $y$ in $x_{0}$, i.e. a cusp at the point $(x_{0},0)$.\par
This property of non differentiability at points for which the speed is zero, implies some difficulties in the calculation of an average profile, particularly in the case of stops. Indeed, if a space-speed profile $v_{1}$ is equal to zero at a point $x_{0}$, and a space-speed profile $v_{2}$ is strictly positive at $x_{0}$, then the sum $v_{1}+v_{2}$ is not a space-speed profile as defined in the Definition \ref{def_space_speed_profiles} since $v_{1}+v_{2}$ is not differentiable at $x_{0}$ but $(v_{1}+v_{2})(x_{0})>0$. Thus, the calculation of an average profile is meaningful only in this two cases:
\begin{itemize}
  \item when all space-speed profiles are strictly positive (no stops) ;
  \item when all space-speed profiles are equal to zero at the same points (i.e. all vehicles stop at the same location).
\end{itemize}
This second case raises the issue of registration of speed profiles that will be discussed in Sect.~\ref{sect_registration}.

\section{Estimation of a space-speed profile from noisy data: A smoothing problem under constraints}
\label{sect_smoothing_problem}

The first step of a functional analysis is to convert the raw data into functional objects that leads to the use of an adapted smoothing procedure. However, the estimation of a space-speed profile from noisy measurements of position and speed is a complex nonparametric regression problem (see for example \citealt{Andrieu2013IV}). Indeed, on the one hand, both the response variable (corresponding to speed) and the explanatory variable (corresponding to vehicle position) are noisy. And on the other hand, the regression function must belong to the space $\mathcal{E}_{SSP}$ defined in Definition \ref{def_space_speed_profiles} and then check its properties, in particular the non differentiability when the speed is zero. To overcome these difficulties, we propose to change to a more suitable study area and start by estimating the function $F$ representing the distance traveled as function of time (study area \texttt{\texttt{distance $\times$ time}} in Fig. \ref{fig_lien_entre_espaces}). Then the new nonparametric model is
\begin{equation}
\label{eq_modele_nonparam_distance_versus_time}
y_{i}=F(t_{i})+\varepsilon_{x,i}, \ \ \ i=1,\ldots,n,
\end{equation}
where $y_{i}$ are noisy observations of the distance traveled, $\varepsilon_{x,i}$ are uncorrelated errors with zero mean and $\sigma_{x}^{2}$ variance, and $F(t)$ is the regression function. This change of study area leads to take into account the two following constraints:
\begin{enumerate}
  \item Use the derivative information, i.e. estimate the regression function $F(t)$ of the model \refeq{eq_modele_nonparam_distance_versus_time} from both noisy observations of $F$ corresponding to position measurements, and noisy observations of its derivative $F'$ corresponding to speed measurements.
  \item A monotonicity constraint since the function $F$ representing the distance traveled as function of time must be increasing.
\end{enumerate}
The consideration of these two constraints in the smoothing step is the subject of the two following subsections. Then, once we have obtained an estimator $\widehat{F}$ of the function $F$, it is easy to deduce by differentiation an estimator $\widehat{F}'$ of the time-speed profile, and finally to deduce an estimator $\widehat{v}_{S}$ of the space-speed profile $v_{S}$ by the transformation $\widehat{v_{S}}=\widehat{F}' \circ \widehat{F}^{-1}$.

\subsection{Smoothing using derivative information}
\label{subsect_smoothing_using_derivative_info}

The first constraint in the nonparametric model \refeq{eq_modele_nonparam_distance_versus_time} is to use the derivative information, i.e. to estimate the regression function $F(t)$ from both noisy measurements of position and speed. Assume that the domain of $F(t)$ is $\mathcal{X}=[0,T]$, where $T$ is a positive real, and $F\in W^{m}[0,T]$ where $W^{m}[0,T]$ is the Sobolev space of order $m$ with $m>1$. Then the use of the derivative information leads to consider the following data model:
\begin{equation}
\label{eq_modele_nonparam_distance_versus_time_use_derivative_info}
\left\{
\begin{array}{l}
y_{i}=F(t_{i})+\varepsilon_{x,i},\ \ \ i=1,\ldots,n \\
v_{i}=F'(t_{i})+\varepsilon_{v,i},\ \ \ i=1,\ldots,n
\end{array}
\right.
\end{equation}
where $y_{i}$ and $v_{i}$ are noisy measurements of distance traveled and speed respectively at each sampling time $t_{i}$, and $\varepsilon_{x,i}$ and $\varepsilon_{v,i}$ are independent zero mean errors with variance $\sigma_{x}^{2}$ and $\sigma_{v}^{2}$ respectively. We also assume that for all $i=1,\ldots,n$, $\varepsilon_{x,i}$ and $\varepsilon_{v,i}$ are independent. Note that we have assumed that the observations $y_{i}$ and $v_{i}$ are obtained at the same times $t_{i}$. Otherwise, a data resampling will lead to this case. \par
The problem of smoothing with derivative information appears in various applications such as economy (\citealt{Hall2007}), molecular biology (\citealt{Calderon2010}) or image analysis (\citealt{Mardia1996}). We propose to solve this problem by using smoothing splines, which have the advantage of requiring the estimation of a single smoothing parameter $\lambda$, contrary to penalized splines used by \cite{Calderon2010}, for which the choice of the number of knots can be difficult. Thus, this problem can be seen as a special case of the general spline smoothing problem (see \citealt{Wahba1990}; \citealt{Wang2011}) and can be solved by using the theory of reproducing kernel Hilbert spaces (see \citealt{Cox1988}). Then, the estimator can be written as a linear combination of basis functions and kernel functions. However, to compute the estimator, it is necessary to choose a norm associated with the design space (in this study, the Sobolev space $W^{m}[0,T]$) that is suitable. Indeed, the expression of the reproducing
kernels used for calculating the estimator depends on the choice of this norm, and can leads to difficulties in the numerical computation (see \citealt{Berlinet2004} for a collection of examples of spaces, norms, and kernels, and \citealt{Andrieu2013thesis}, Chap. 4 for more details). So, we propose to use the theory of thin-plate spline which leads to similar results for the form of the estimator (see \citealt{Duchon1977}; \citealt{Meinguet1979}; \citealt{Wahba1980}), but where the reproducing kernel is replaced by a semi-kernel with a simpler form that greatly simplifies the computation of the estimator. \\

The model \refeq{eq_modele_nonparam_distance_versus_time_use_derivative_info} is a particular case of the general spline smoothing model defined in \cite{Wahba1990}, since it can be rewritten as follows:
\begin{equation}
\label{eq_modele_general_spline_de_lissage_use_derivative_info}
y_{j}=\mathcal{L}_{j}F+\varepsilon_{j} \ , \ \ j=1,\ldots,2n,
\end{equation}
where
\begin{itemize}
  \item the observations $y_{j}$ are defined by: \\
    $\left\{
     \begin{array}{l}
      y_{j}=y_{i} \ \ \text{with}\ \ i=j \ \ \text{for}\ \ j=1,\ldots,n \\
      y_{j}=v_{i} \ \ \text{with}\ \ i=j-n \ \ \text{for}\ \ j=n+1,\ldots,2n
     \end{array}
   \right.$,\\
  \item the bounded linear functionals $\mathcal{L}_{j}$ on $W^{m}[0,T]$ are defined by: \\
    $\left\{
     \begin{array}{l}
      \mathcal{L}_{j}F=F(t_{i}) \ \ \text{with}\ \ i=j \ \ \text{for}\ \ j=1,\ldots,n \\
      \mathcal{L}_{j}F=F'(t_{i}) \ \ \text{with}\ \ i=j-n \ \ \text{for}\ \ j=n+1,\ldots,2n
     \end{array}
    \right.$,\\
  \item the errors $\varepsilon_{j}$ are defined by: \\
    $\left\{
     \begin{array}{l}
      \varepsilon_{j}=\varepsilon_{x,i} \ \ \text{avec}\ \ i=j \ \ \text{pour}\ \ j=1,\ldots,n \\
      \varepsilon_{j}=\varepsilon_{v,i} \ \ \text{avec}\ \ i=j-n \ \ \text{pour}\ \ j=n+1,\ldots,2n
     \end{array}
    \right.$.
\end{itemize}
A useful result state that the operator $L$ defined by $Lf=\frac{\partial^{k}f}{\partial^{\alpha_{1}}x_{1}\ldots\partial^{\alpha_{d}}x_{d}}$ for $\alpha_{1}+\ldots+\alpha_{d}=k$ ($k,\alpha_{1},\ldots,\alpha_{d}\in\mathds{N}$) is a continuous linear form if and only if $2m-2k-d>0$ (see \citealt{Berlinet2004}, Th. 133;  \citealt{Wahba1980}). So, we can deduce that linear functionals in the model \refeq{eq_modele_general_spline_de_lissage_use_derivative_info} are bounded on $W^{m}[0,T]$ if $m>1$ since $d=1$, $k=0$ for $j=1,\ldots,n$ and $k=1$ for $j=n+1,\ldots,2n$. Then, an estimator of $F(t)$ is the minimizer of the following penalized least squares criterion in $W^{m}[0,T]$:
\begin{equation}
\label{eq_application_ssp_critere_modele_spline_de_lissage_use_derivative_info}
\frac{1}{2n} \{\sigma_{x}^{-2} \sum_{i=1}^{n} (y_{i}-F(t_{i}))^{2} + \sigma_{v}^{-2} \sum_{i=1}^{n} (v_{i}-F'(t_{i}))^{2} \}  + \lambda \int_{0}^{T}(F^{(m)}(t))^{2}dt.
\end{equation}

A solution to a more general minimization problem extended to dimension $d\geq 1$ of which the minimization problem \refeq{eq_application_ssp_critere_modele_spline_de_lissage_use_derivative_info} is a special case with $d=1$, is given in \cite{Wahba1980}. Before stating the result, we introduce the $m$ polynomials functions $\phi_{1},\ldots,\phi_{m}$ as a basis of the null space of the penalty functional $J_{m}(f)=\int_{0}^{T}(f^{(m)}(t))^{2}dt$, i.e. $\mathcal{H}_{0}=\{f : J_{m}(f)=0\}$. Thus, provided that the two hypothesis on the linear functionals ($L_{1},\ldots,L_{n}$ linearly independent continuous linear functionals, and $L_{k}\sum_{\nu=1}^{p}a_{\nu}\phi_{\nu}=0$ implies that all the $a_{\nu}$ are $0$) are satisfied, which it's the case if the sampling points $t_{1},\ldots,t_{n}$ are distincts, we can deduce that the minimizer of \refeq{eq_application_ssp_critere_modele_spline_de_lissage_use_derivative_info} can be written as:
\begin{equation}
\label{eq_application_ssp_solution_critere_modele_general_thin_plate_spline_use_derivative_info}
\widehat{F}_{\lambda}(t)=\sum_{\nu=1}^{m} d_{\nu} \phi_{\nu}(t) + \sum_{i=1}^{n} c_{i}E_{m}(t_{i},t) + \sum_{i=1}^{n} c'_{i}\frac{\partial}{\partial s}E_{m}(s,t)|_{s=t_{i}},
\end{equation}
where
\begin{eqnarray*}
E_{m}(s,t)    & = &   \theta_{m,1}|s-t|^{2m-1}, \nonumber \\
\theta_{m,1}  & = &   \frac{\Gamma(1/2-m)}{2^{2m}\pi^{1/2}(m-1)!}. \nonumber \\
\end{eqnarray*}
The coefficients $c=(c_{1},\ldots,c_{n},c'_{1},\ldots,c'_{n})^{T}$ and $d=(d_{1},\ldots,d_{m})^{T}$ are solutions of the linear system
\begin{eqnarray}
(K+n\lambda W^{-1}) c +  T d  & = &  y,  \label{eq_modele_general_thin_plate_spline_syst_coeff_eq1}\\
T^{T} c     & = &  0, \label{eq_modele_general_thin_plate_spline_syst_coeff_eq2}
\end{eqnarray}
where
\begin{eqnarray*}
T             & = & {\{\mathcal{L}_{j}\phi_{\nu}\}  \scriptsize  \begin{matrix} 2n & \mkern-10mu m  \\j=1 & \mkern-5mu \nu=1 \end{matrix}}, \nonumber \\
K       & = &   \{\mathcal{L}_{j(s)}\mathcal{L}_{k(t)}E_{m}(s,t)\}_{j,k=1}^{2n}, \nonumber \\
W             & = &    diag(\sigma_{x}^{-2},\ldots,\sigma_{x}^{-2},\sigma_{v}^{-2},\ldots,\sigma_{v}^{-2}). \nonumber \\
\end{eqnarray*}
The solution $\widehat{F}_{\lambda}(t)$ is then a polynomial spline of order $m$ with knots at the sampling time $t_{1},\ldots,t_{n}$. Note that the bivariate function $E_{m}(s,t)$, called semi-kernel, acts like a reproducing kernel in this approach, but its simple form is more appropriate for computational aspect and specifically for solving the linear system defined by Eqs. \refeq{eq_modele_general_thin_plate_spline_syst_coeff_eq1} and \refeq{eq_modele_general_thin_plate_spline_syst_coeff_eq2}. \\

The error variances $\sigma_{x}^{2}$ and $\sigma_{v}^{2}$ are usually unknown in practice. In general, we use an estimator of the error variance corresponding to the criterion used for the selection of the smoothing parameter $\lambda$. Three scores are commonly used:
\begin{itemize}
  \item the UBR score ("Unbiased Risk") which is an extension of the Mallow's $C_{p}$ criterion ;
  \item the GCV score ("Generalized Cross-Validation") which is a weighted version of the standard cross-validation ;
  \item the GML score ("Generalized Maximum Likelihood") based on a Bayesian model, and that required a normality assumption on the errors.
\end{itemize}
The selection of the smoothing parameter results from the minimization of one of these criteria, and the error variances estimates depend on the smoothing parameter obtained, and therefore on the criterion chosen (see \citealt{Gu2002}). Thus, if in a first time, we consider only the position measurements, i.e. the data model \refeq{eq_modele_nonparam_distance_versus_time}, and if we denote $A(\lambda)$ the hat matrix defined by
\begin{equation}
\label{eq_hat_matrix}
(\widehat{F}_{\lambda}(t_{1}),\ldots,\widehat{F}_{\lambda}(t_{n}))^{T} = A(\lambda)y,
\end{equation}
where $\widehat{F}_{\lambda}$ is the smoothing spline estimate of $F$ for the smoothing parameter $\lambda$ (that is a polynomial spline of order $2m$), then the variance estimate of $\sigma_{x}^{2}$ associated with the GCV criterion is
\begin{equation}
\label{eq_application_estimation_variance_critere_gcv}
\widehat{\sigma}^{2}_{gcv}=\frac{y^{T}(I-A(\lambda_{v}))^{2}y}{tr(I-A(\lambda_{v}))},
\end{equation}
and the variance estimate associated with the GLM criterion is
\begin{equation}
\label{eq_application_estimation_variance_critere_glm}
\widehat{\sigma}^{2}_{gml}=\frac{y^{T}(I-A(\lambda_{m}))y}{n-m}.
\end{equation}
Similarly, in a second time, we consider only the speed measurements and calculate the smoothing spline estimate of $F'$ for the smoothing parameter $\lambda$, and then deduce an estimate of the variance error $\sigma_{v}^{2}$.

\subsection{Smoothing under monotonicity constraint}
\label{subsect_smoothing_under_monotonicity_constraint}

The second constraint in the nonparametric model \refeq{eq_modele_nonparam_distance_versus_time} is a monotonicity constraint since the function $F$ representing the distance traveled as a function of time must be increasing. Various methods of smoothing under monotonicity constraint have been developed. The main approaches are based on kernel smoothers and splines. An overview of these methods can be found in \cite{Delecroix2000}. Among the main methods, we can cite the isotonic regression introduced by \cite{Brunk1955}, the monotone splines (for example, Ramsay introduces the I-splines basis in \citealt{Ramsay1988} for monotone regression splines) or the projection methods (e.g. \citealt{Delecroix1996} or \citealt{Mammen2001}). \par
In a previous study (\citealt{Andrieu2012Jds}), the method of homeomorphic splines developed by \cite{Bigot2010} have been tested. However, if the monotonization step presented good results, we had difficulties in the implementation of the derivative. So, we propose to use a method developed by \cite{Ramsay1998} which has the advantage of being relatively simple to implement. The principle of this method is to transform the constrained smoothing problem to an unconstrained one. A monotone function has a positive first derivative. So the main idea is that any strictly monotonic function $f$ satisfies the following differential equation:
\begin{equation}
\label{eq_monotone_function_representation_by_eq_diff}
D^{2}f=w Df,
\end{equation}
where $Df$ and $D^{2}f$ are respectively the first and second derivative of the function $f$, and $w$ is an unconstrained function. Thus, any strictly monotonic function $f$ can be written as following (as solution of the Eq. \refeq{eq_monotone_function_representation_by_eq_diff}):
\begin{equation}
\label{eq_monotone_function_solution}
f(t)=\beta_{0}+\beta_{1} \int_{0}^{t} exp[ \int_{0}^{u} w(v) dv] du,
\end{equation}
where $\beta_{0}$ and $\beta_{1}$ are arbitrary constants such that $f(0)=\beta_{0}$ and $f'(0)=\beta_{1}$. Then, the problem is to estimate the coefficients $\beta_{0}$ and $\beta_{1}$ and the unconstrained function $w$ by minimizing the following criterion:
\begin{equation}
\label{eq_monotone_function_criterion}
\sum_{i=1}^{n} (y_{i}-\beta_{0}-\beta_{1}h(t_{i}))^{2} + \lambda \int_{0}^{T} (w^{m}(t))^{2} dt,
\end{equation}
where
\begin{equation}
\label{eq_monotone_function_criterion_suite}
h(t)=\int_{0}^{t} exp[\int_{0}^{v}w(v)dv] du.
\end{equation}
The unconstrained function $w$ is computed using an appropriate basis expansion (e.g. B-splines) and the coefficients $\beta_{0}$ and $\beta_{1}$ are estimated by numerical algorithms. However, due to a numerical optimization of the criterion \refeq{eq_monotone_function_criterion}, monotone smoothing spline involves considerably more computation than the usual smoothing spline process.\\

Thus, the monotonicity constraint is consider in a second smoothing step by applying the method of Ramsay described above, to the estimated values $\widehat{F}_{\lambda}(t_{i})$ obtained at the first smoothing step for which we have used the speed measurements (Section \ref{subsect_smoothing_using_derivative_info}). Therefore, this second smoothing step can be seen as a monotonization step and then is similar to the projection step in projection methods (see \citealt{Mammen2001}). The new monotone estimator of $F(t)$ is then denoted $\widehat{F}_{mc}$ (with "mc" for monotonicity constraint), and finally we deduce an estimator $\widehat{v}_{S}$ of the space-speed profile $v_{S}$ by the transformation $\widehat{v_{S}}=\widehat{F}_{mc}^{'} \circ \widehat{F}_{mc}^{-1}$.

\subsection{Simulation study}
\label{subsect_simulation_studies}

We propose to illustrate the performance of our smoothing procedure on three simulated examples. For each example, we study the monotone estimator $\widehat{F}_{mc}$ obtained after the two steps of the smoothing procedure (use of derivative information, and then monotonization), the derivative $\widehat{F}'_{mc}$ of this estimator, and the composite function $\widehat{F}_{mc}^{'} \circ \widehat{F}_{mc}^{-1}$. \par
\begin{figure}[!h]
\vspace{2 pt}
\centerline{
\begin{tabular}{ccc}
\includegraphics[width=4cm]{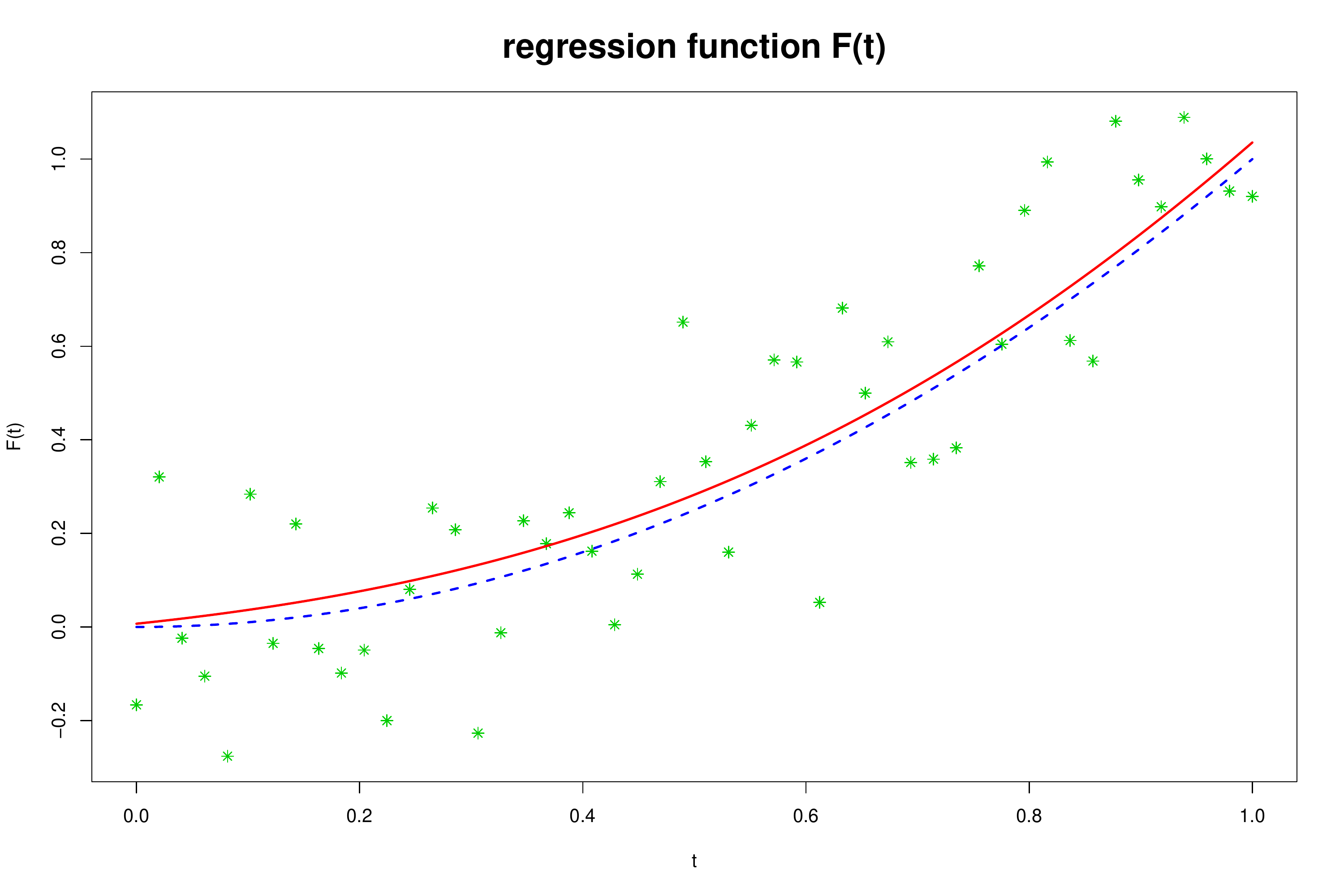} &
\includegraphics[width=4cm]{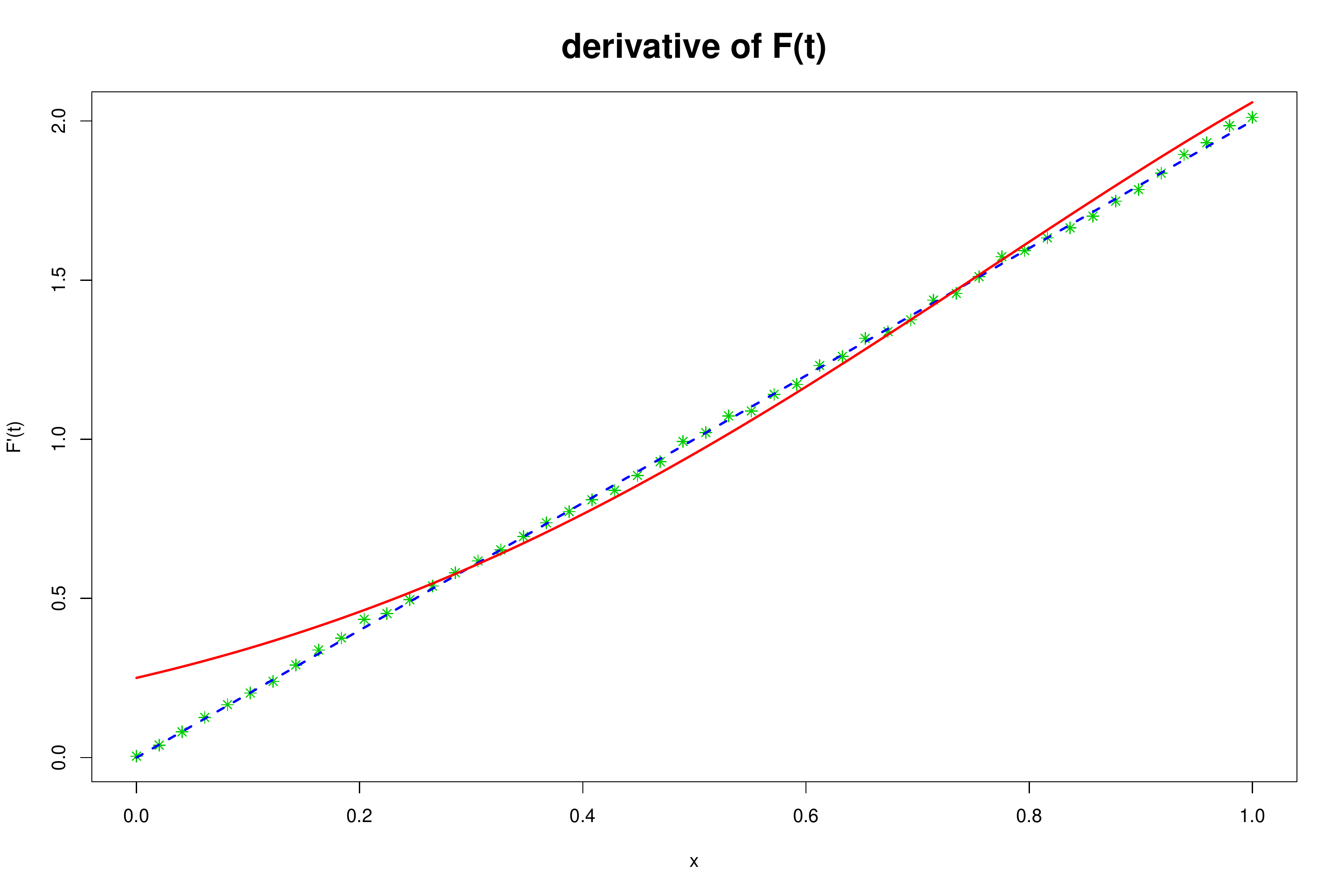} &
\includegraphics[width=4cm]{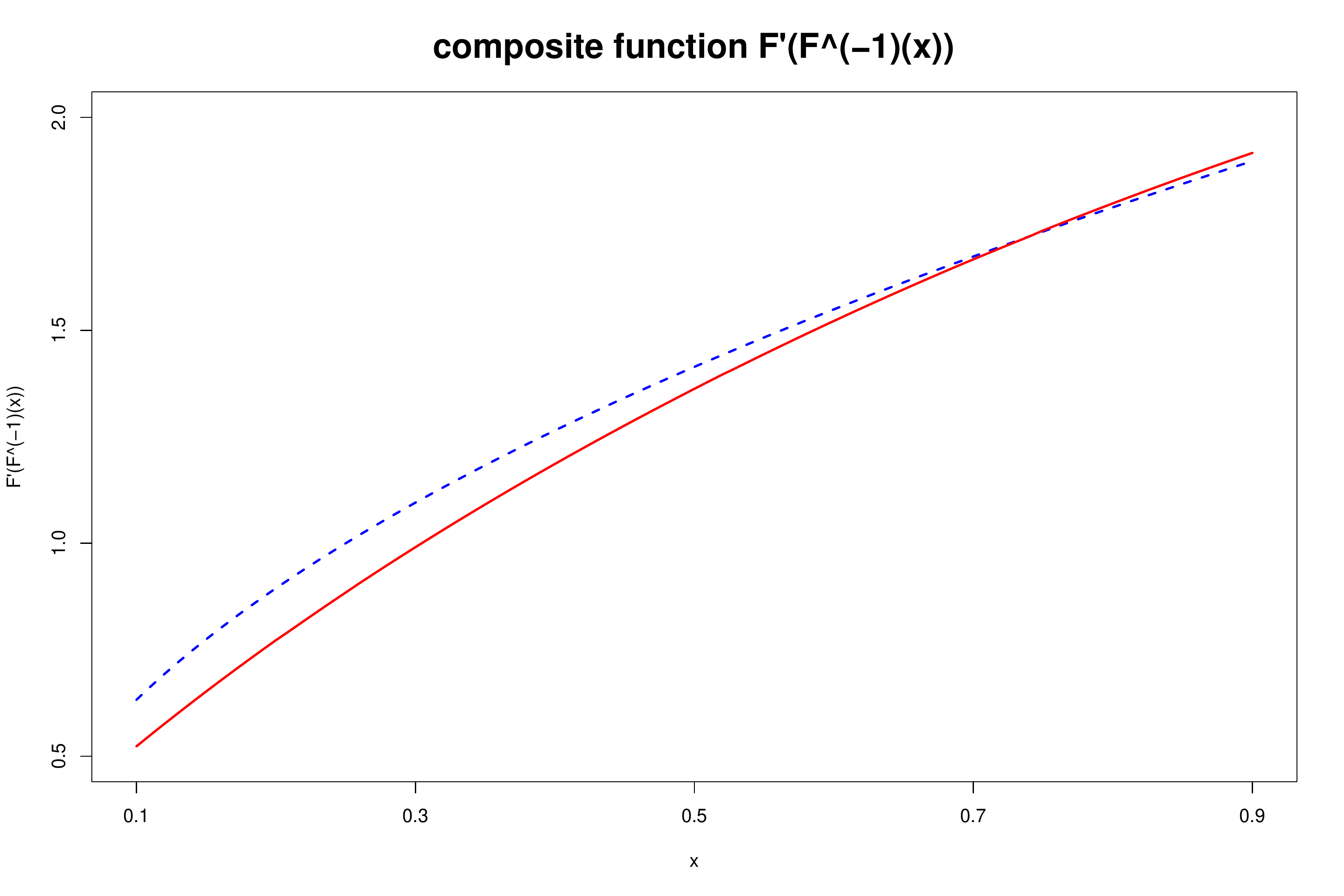} \\
\includegraphics[width=4cm]{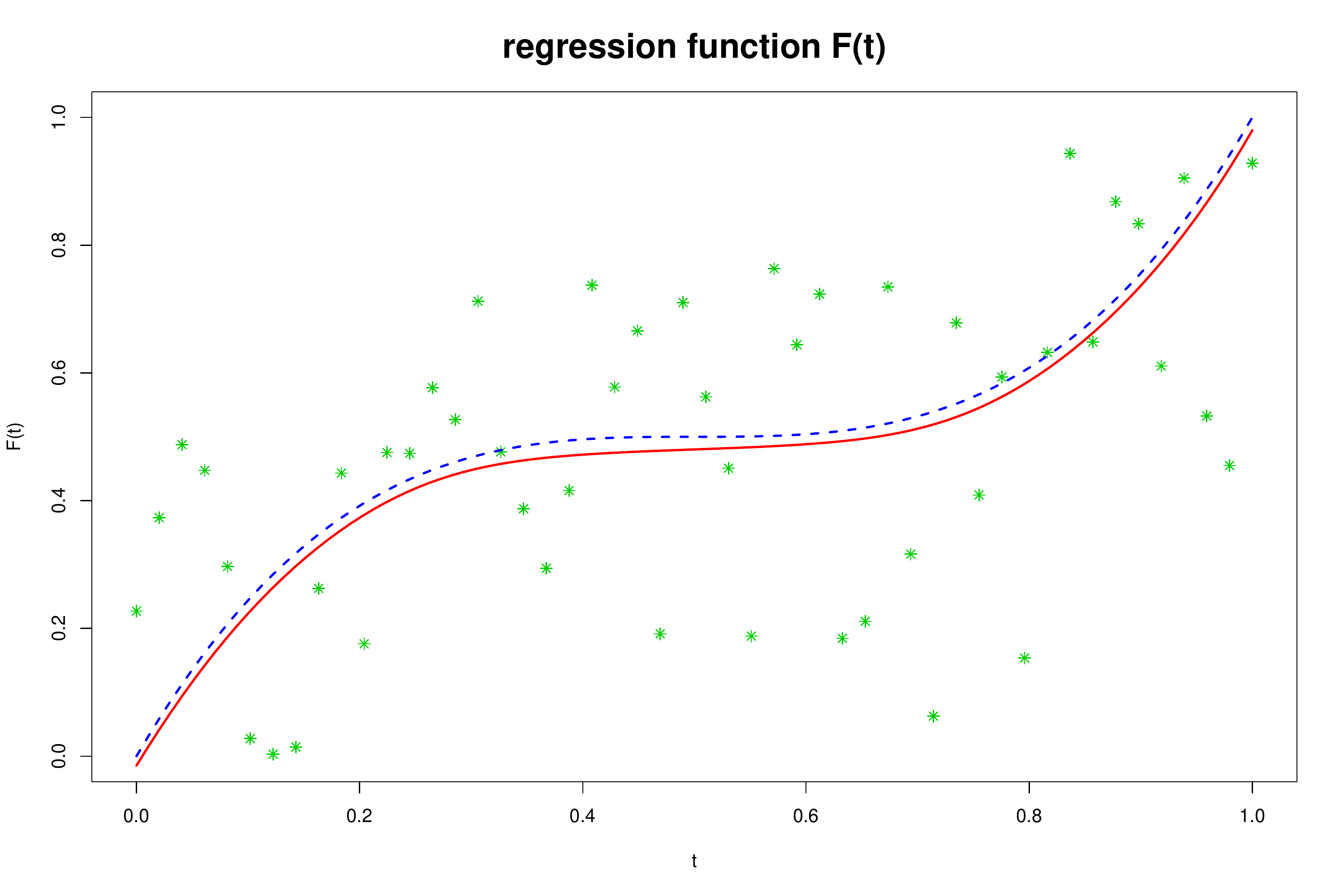} &
\includegraphics[width=4cm]{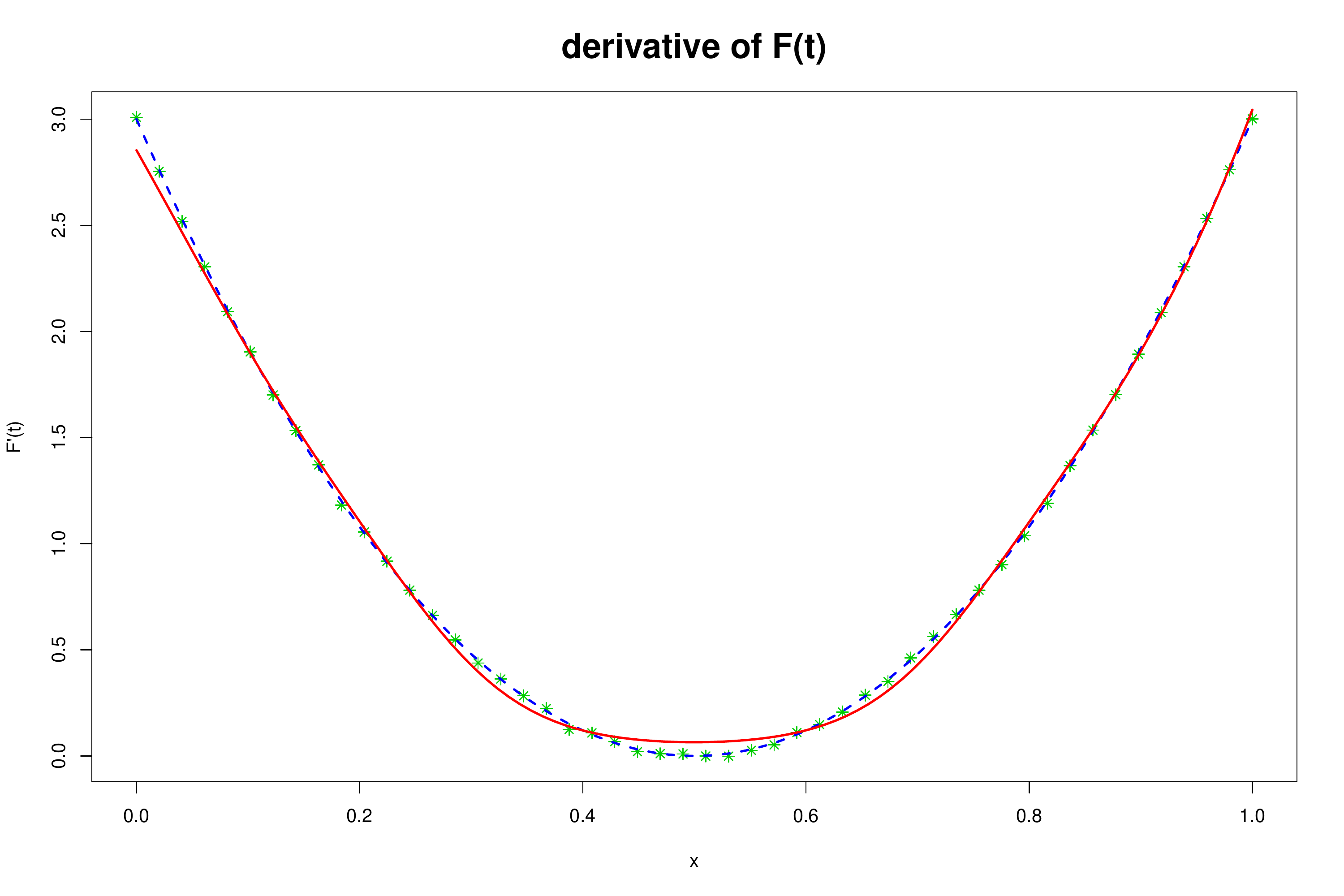} &
\includegraphics[width=4cm]{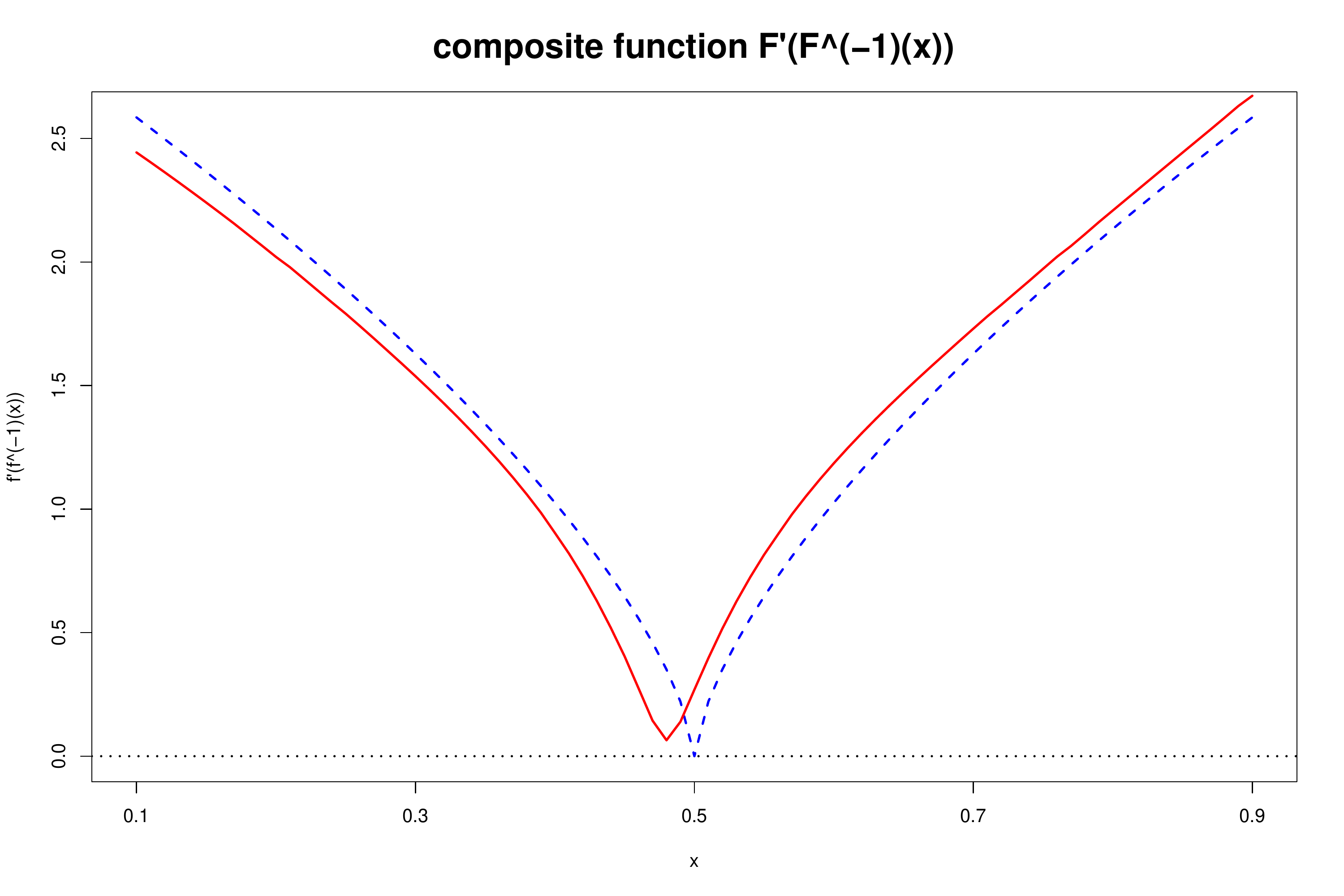} \\
\includegraphics[width=4cm]{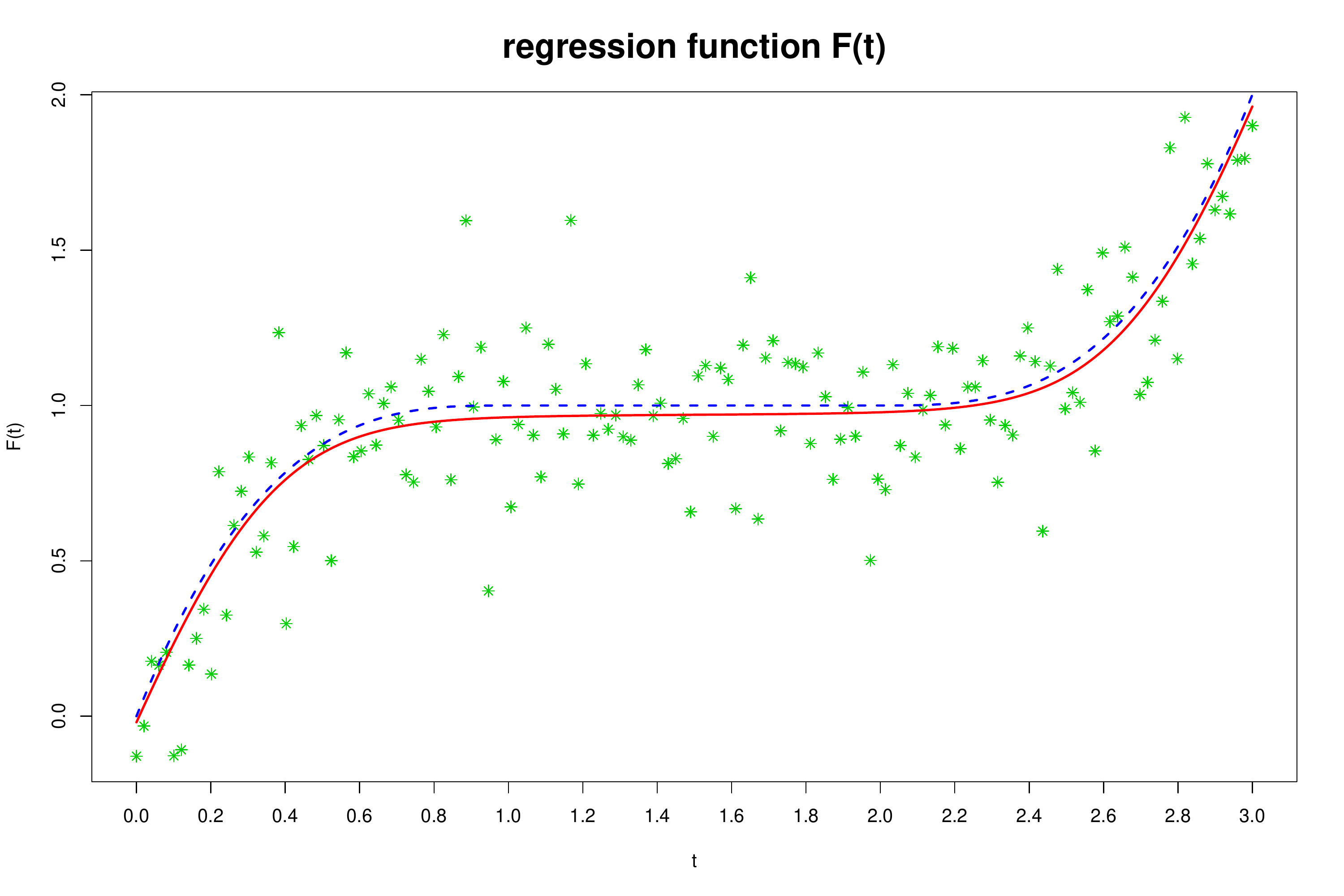} &
\includegraphics[width=4cm]{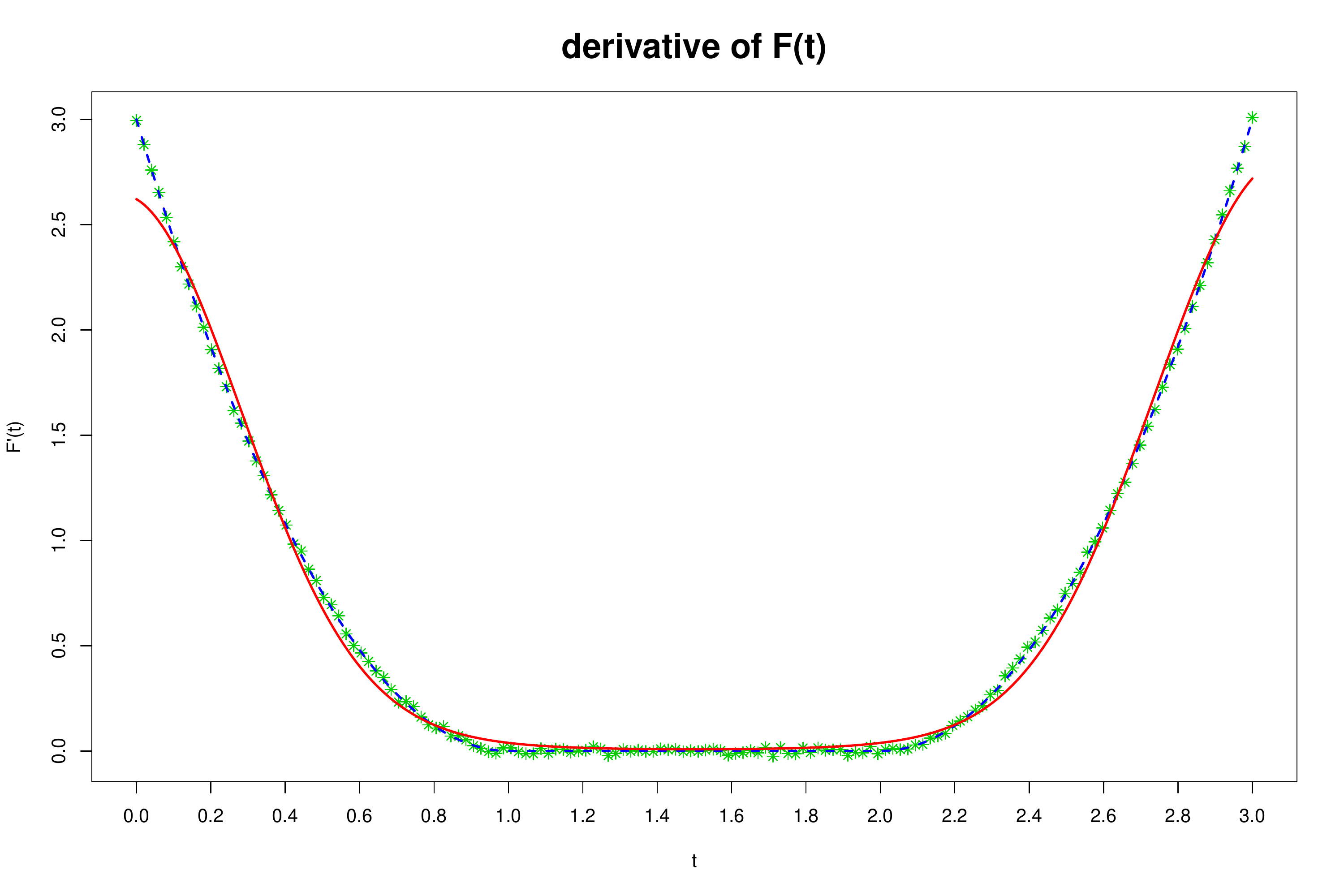} &
\includegraphics[width=4cm]{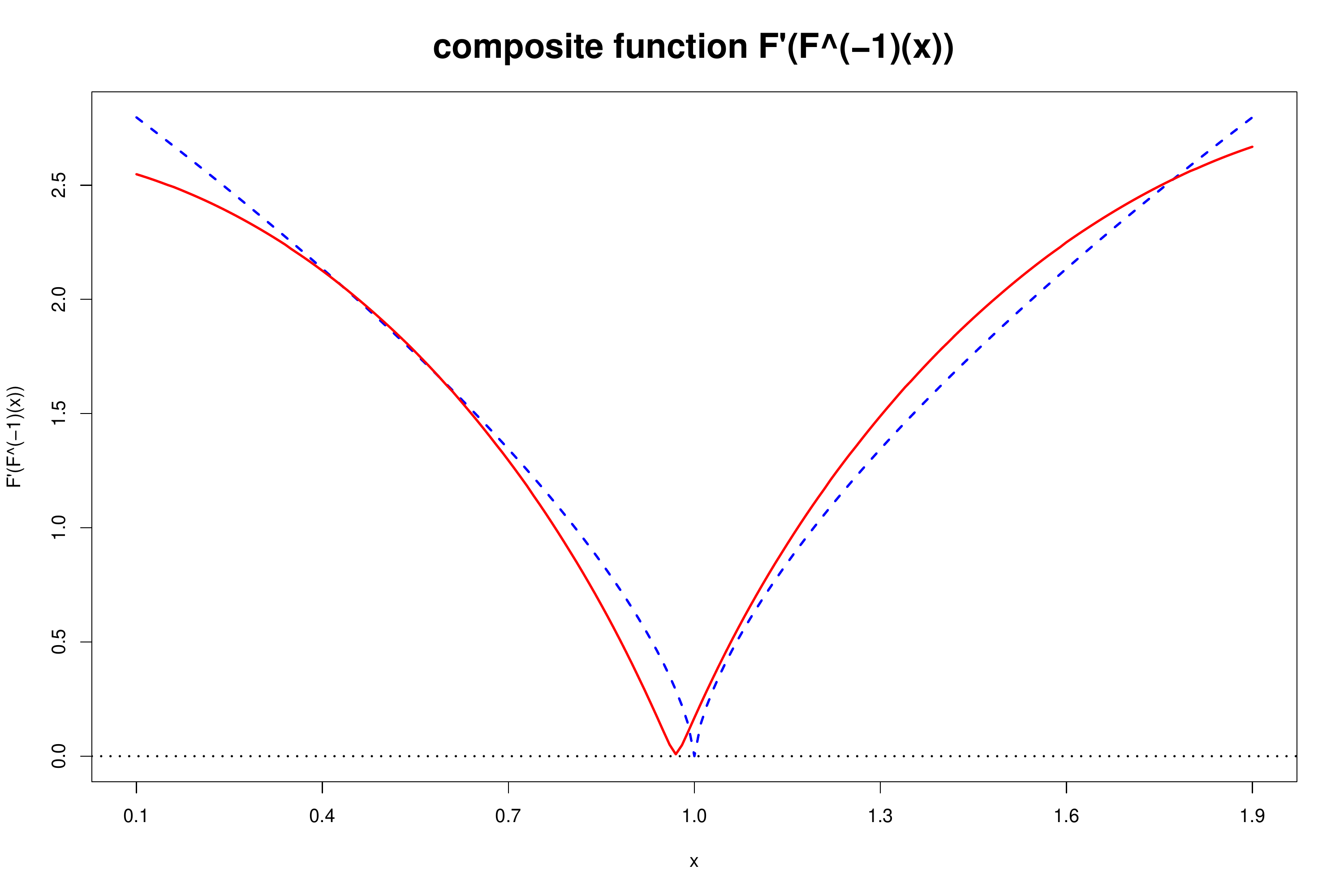} \\
\end{tabular}
}
\caption{Estimators of $F$, $F'$ and $F' \circ F^{-1}$ (left to right) for the three examples \refeq{eq_simulation_study_f1}-\refeq{eq_simulation_study_f3} (top to bottom). The unknown regression function $F$, its derivative $F'$ and the composite function $F' \circ F^{-1}$ are dashed blue lines. The noisy data are green points.  The estimators $\widehat{F}_{mc}$, $\widehat{F}'_{mc}$ and $\widehat{F}_{mc}^{'} \circ \widehat{F}_{mc}^{-1}$ are solid red lines.}
\label{fig_simulation_studies_estimators}
\end{figure}

We investigate the regression model \refeq{eq_modele_nonparam_distance_versus_time_use_derivative_info} with a fixed design made up of $n=50$ points evenly distributed on $[0,1]$ for the first and second example, and $n=150$ points evenly distributed on $[0,3]$ for the third example. The errors term $\varepsilon_{x,i}$ and $\varepsilon_{v,i}$ were simulated from centered gaussian distributions with $\sigma_{x}=0.2$ and $\sigma_{v}=0.01$. This choice of a smaller noise for the derivative of $F$ than for $F$ is motivated by the fact that speed measurements are generally more accurate than position measurements. The increasing regression functions chosen in these three examples are:
\begin{eqnarray}
F_{1}(t) & = & t^{2}  \ \ \text{with}\ \ t \in [0,1], \label{eq_simulation_study_f1} \\
F_{2}(t) & = & \frac{1}{2}(2t-1)^{3}+\frac{1}{2} \ \ \text{with}\ \ t \in [0,1], \label{eq_simulation_study_f2} \\
F_{3}(t) & = & \left\{
    \begin{array}{ll}
        (t-1)^{3}+1 & \mbox{si } t \leq 1 \\
        1 & \mbox{si } 1 \leq t \leq 2 \\
        (t-2)^{3}+1 & \mbox{si } t \geq 2
    \end{array}
\right. \ \ \text{with}\ \ t \in [0,3], \label{eq_simulation_study_f3}
\end{eqnarray}
and correspond respectively to a convex function, a function with a small plateau (inflection point), and a function with a large plateau ($F_{3}$ is an extension of $F_{2}$ with a plateau over $[1,2]$). The different functions, their derivatives, and the composite functions $F_{i}' \circ F_{i}^{-1}$, $i=1,2,3$, are displayed Fig. \ref{fig_simulation_studies_estimators} (dashed blue lines). \par
The proposed estimator $\widehat{F}_{mc}$ of the the unknown regression function $F$ for the three examples \refeq{eq_simulation_study_f1}-\refeq{eq_simulation_study_f3} is the solid red line (left column in Fig. \ref{fig_simulation_studies_estimators}). As described in the previous section, this estimator was calculated in two steps. The first step corresponding to the construction of the estimator \refeq{eq_application_ssp_solution_critere_modele_general_thin_plate_spline_use_derivative_info} that take into account the derivative information, was computed with the function \texttt{ssr} in the R (\citealt{Rsoftware2013}) package \texttt{assist} (\citealt{Wang2004}). In the three examples, we have chosen $m=3$ (i.e. a quintic spline) for the order of the Sobolev space, and the GML criterion was used for the selection of the smoothing parameter. The second step corresponding to the monotonization of the estimator \refeq{eq_application_ssp_solution_critere_modele_general_thin_plate_spline_use_derivative_info} was computed with the function \texttt{smooth.monotone} in the R package \texttt{fda}. We have chosen $m=3$ for the degree of the penalty in Eq. \refeq{eq_monotone_function_criterion}, and the smoothing parameter was chosen by trial and error. \par
The derivative $\widehat{F}'_{mc}$ of the estimator is represented in the middle column of the Fig. \ref{fig_simulation_studies_estimators}, and the composite function $\widehat{F}_{mc}^{'} \circ \widehat{F}_{mc}^{-1}$ is in the right column (red lines). The computation of a point $t_{0}=F^{-1}(y_{0})$ was made by the computation of the root (i.e. zero) of the function $F(t)-y_{0}$ (use of the R function \texttt{uniroot}). Due to computational problems of the inverse function at the edge of the interval, the composite function $F' \circ F^{-1}$ and its estimator was computed over [0.1,0.9] for the first and second example \refeq{eq_simulation_study_f1}-\refeq{eq_simulation_study_f2}, and over [0.1,1.9] for the third example \refeq{eq_simulation_study_f3}. \par

~\\

\begin{figure}[!h]
\vspace{2 pt}
\centerline{
\begin{tabular}{ccc}
\includegraphics[width=4cm]{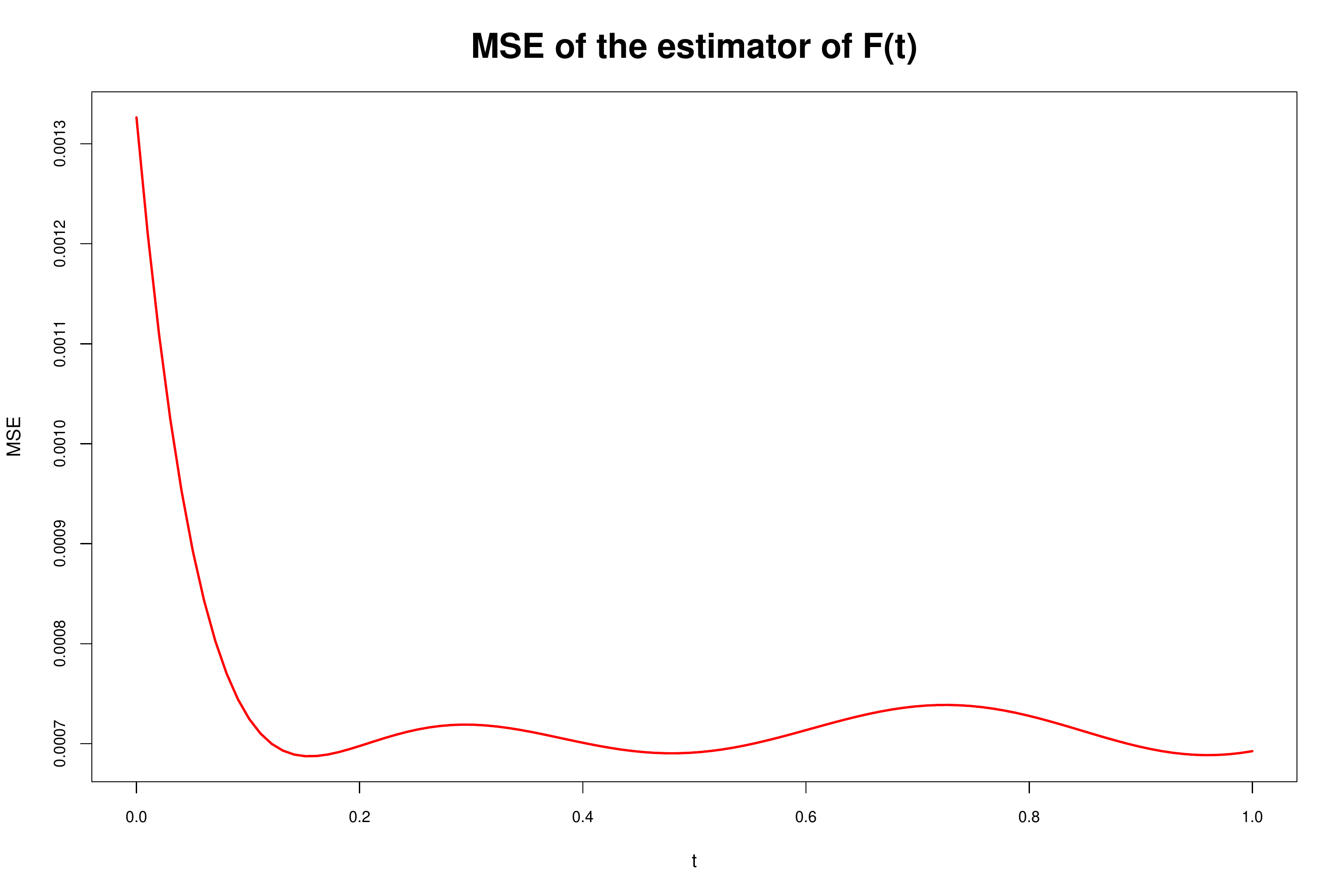} &
\includegraphics[width=4cm]{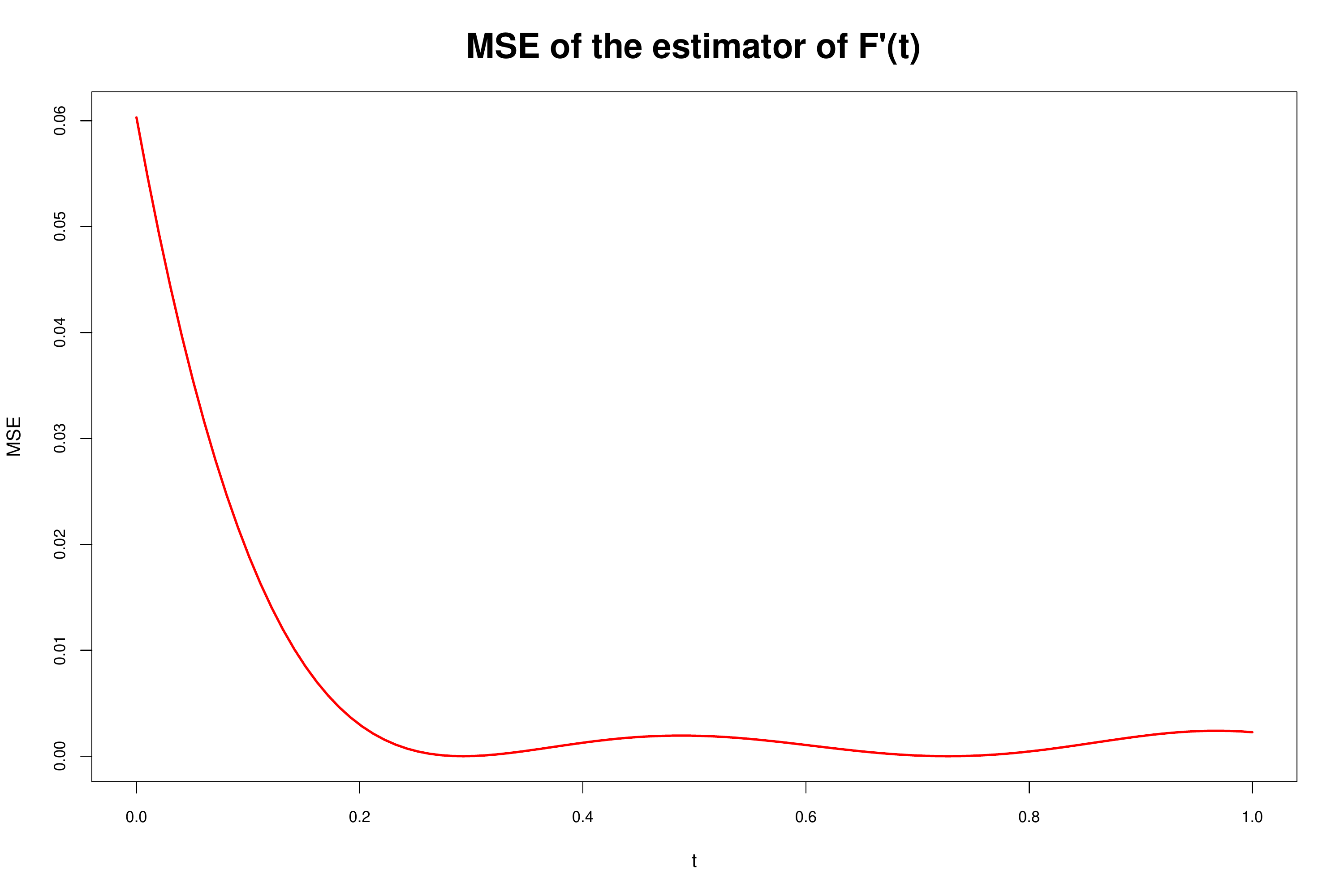} &
\includegraphics[width=4cm]{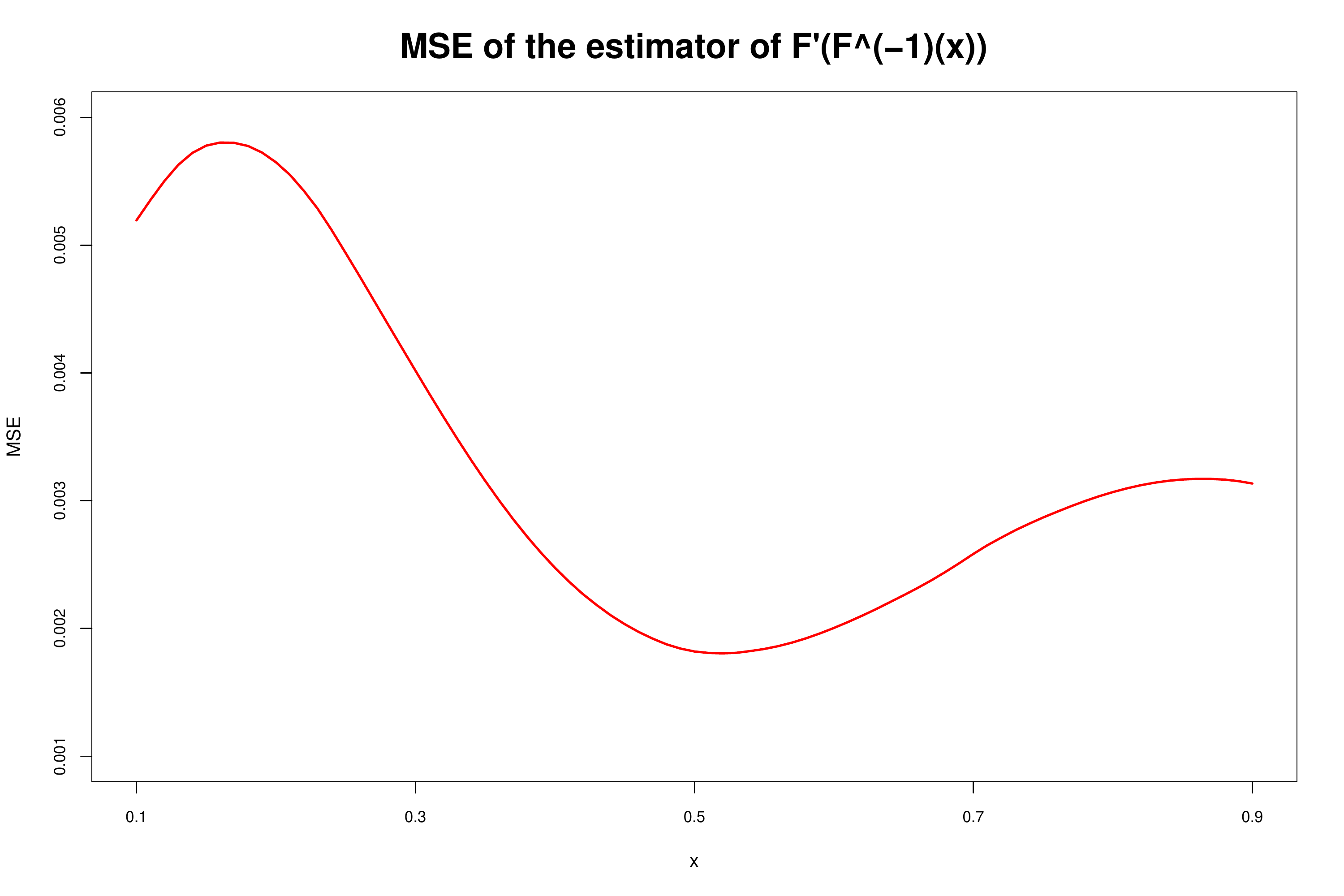} \\
\includegraphics[width=4cm]{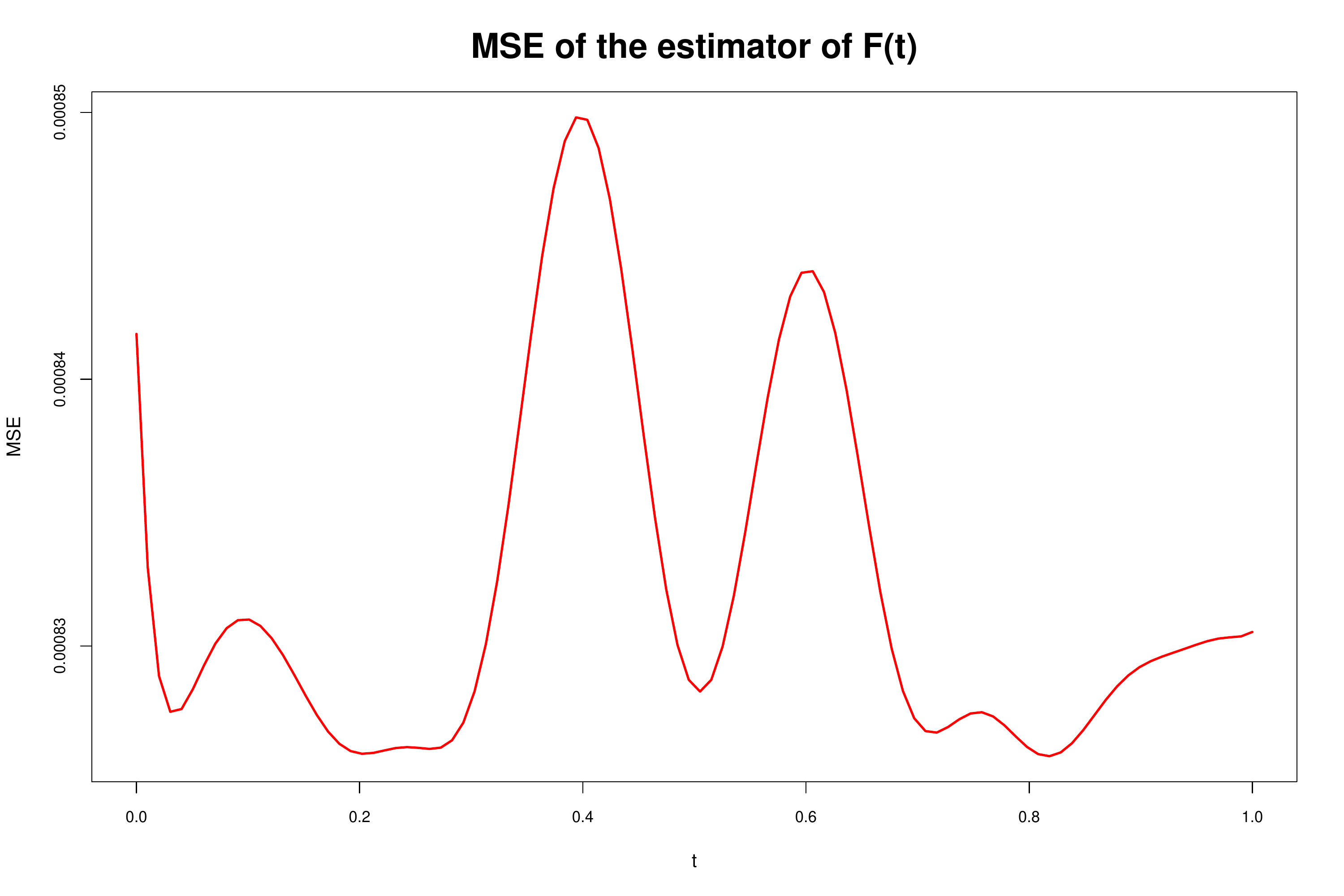} &
\includegraphics[width=4cm]{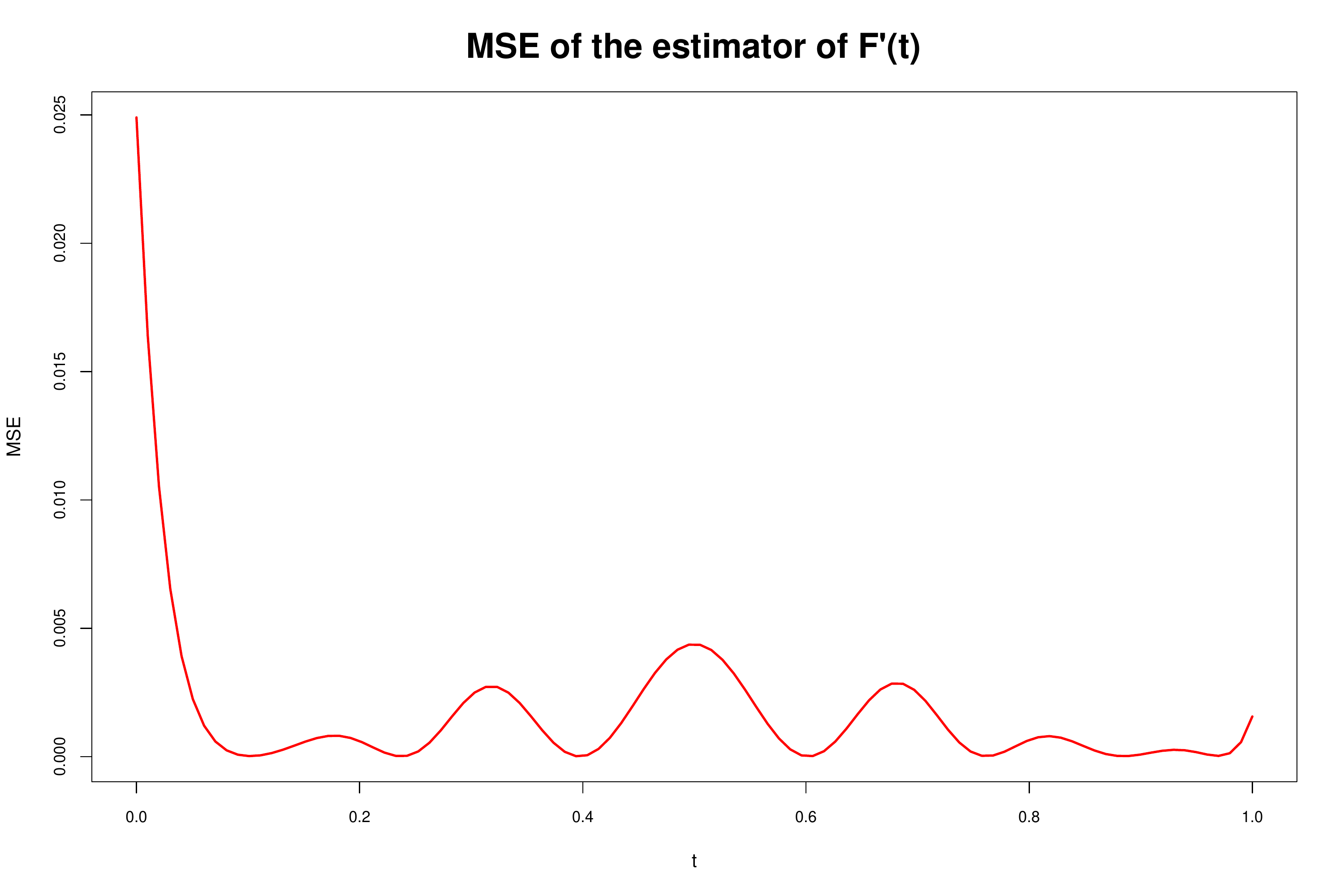} &
\includegraphics[width=4cm]{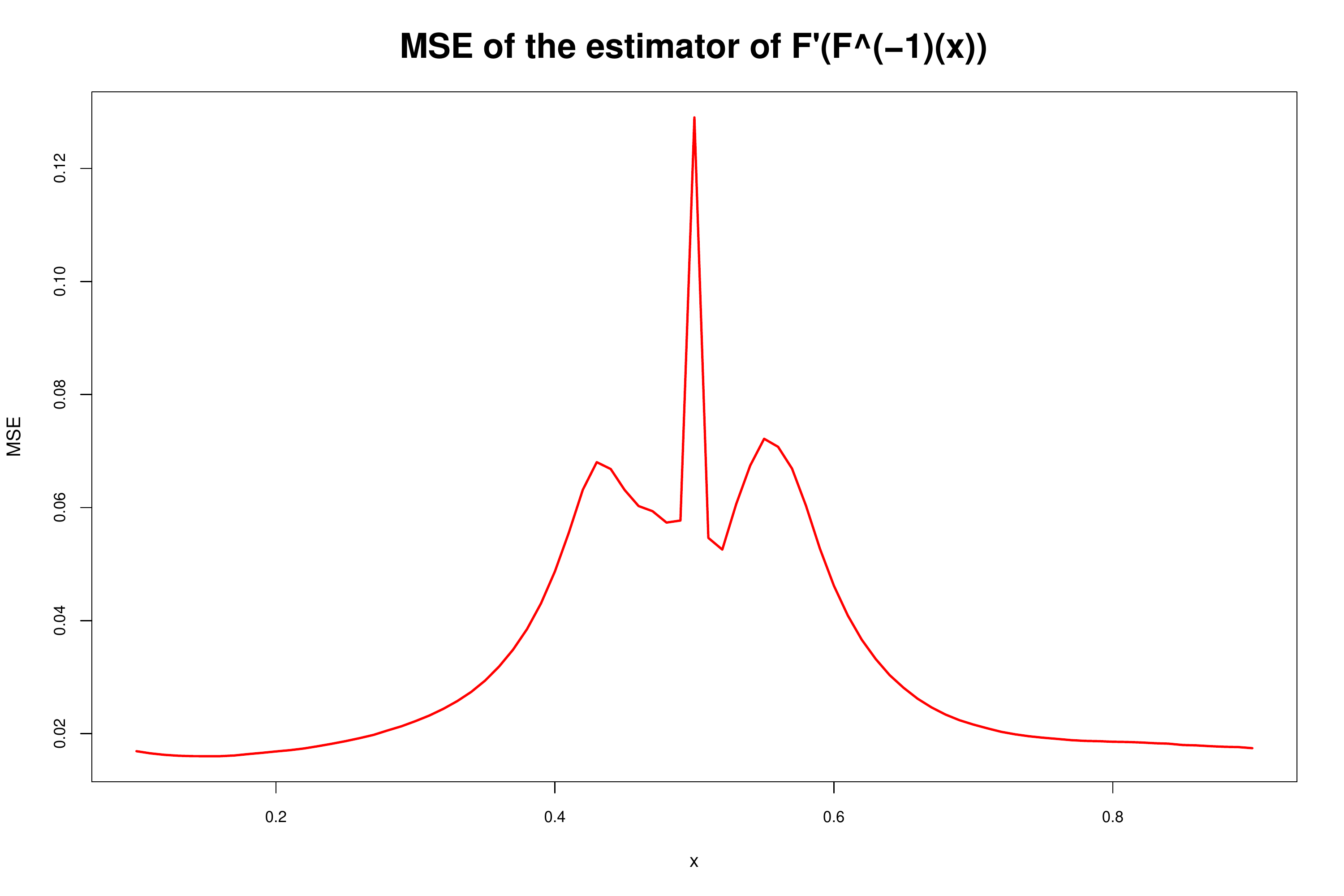} \\
\includegraphics[width=4cm]{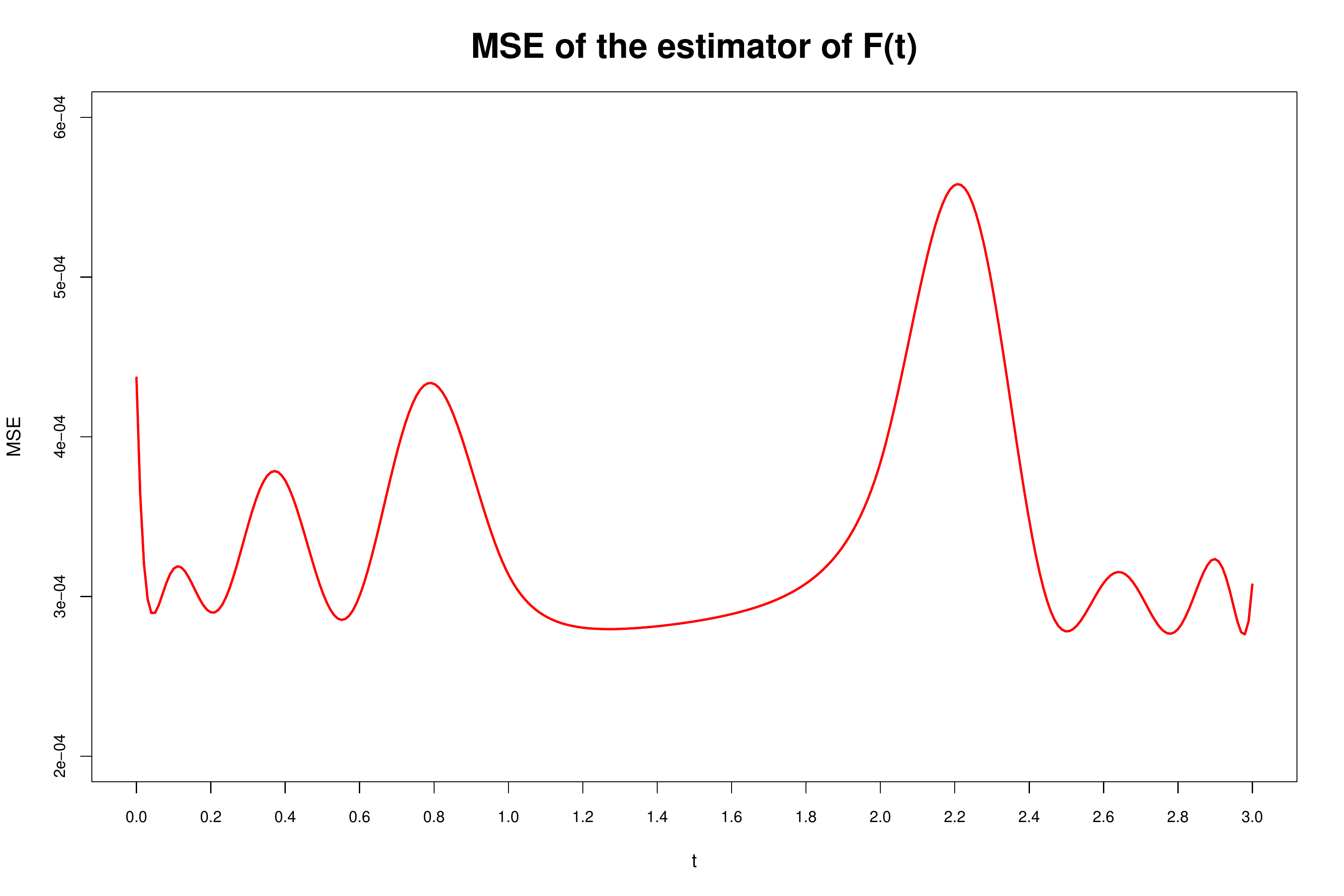} &
\includegraphics[width=4cm]{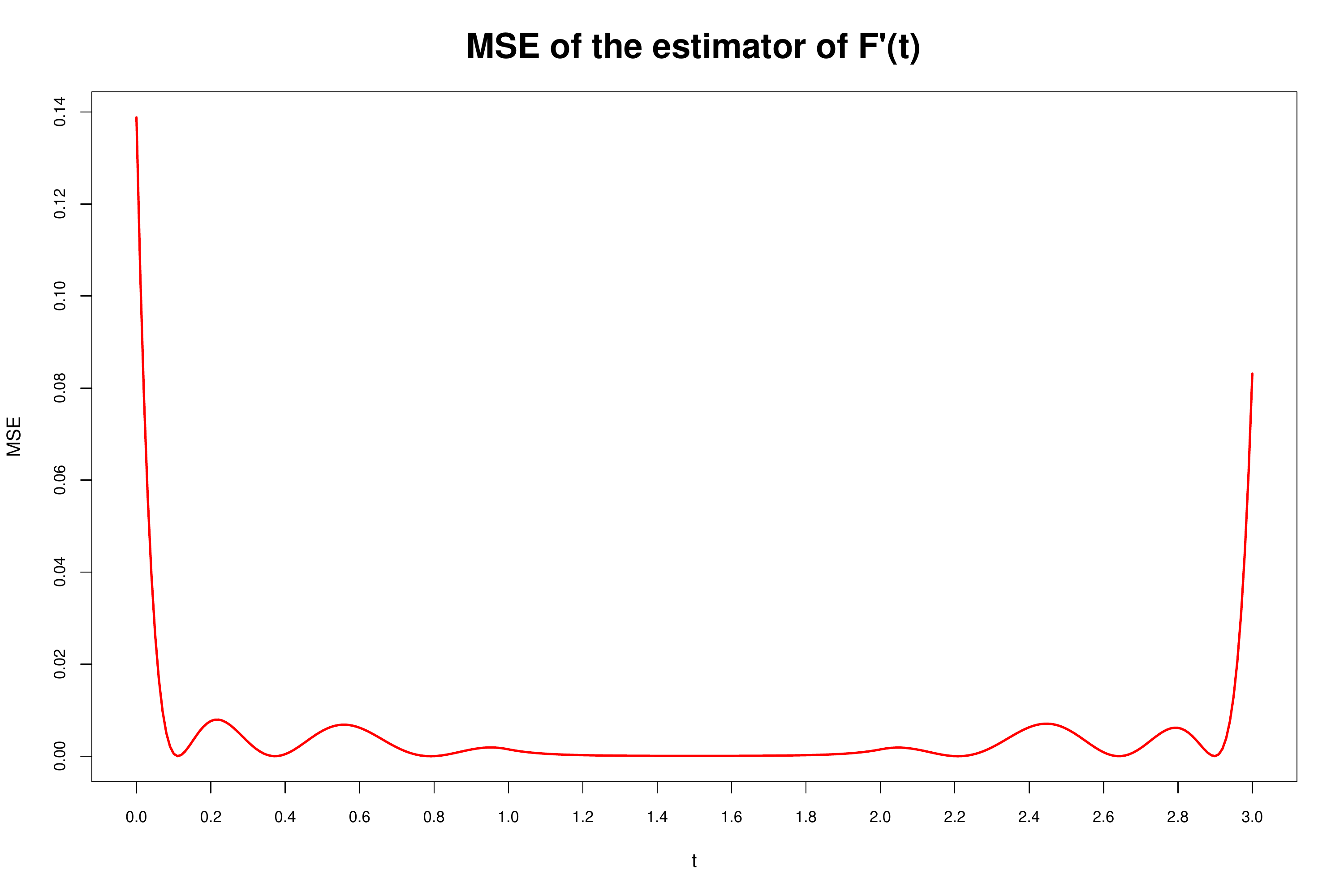} &
\includegraphics[width=4cm]{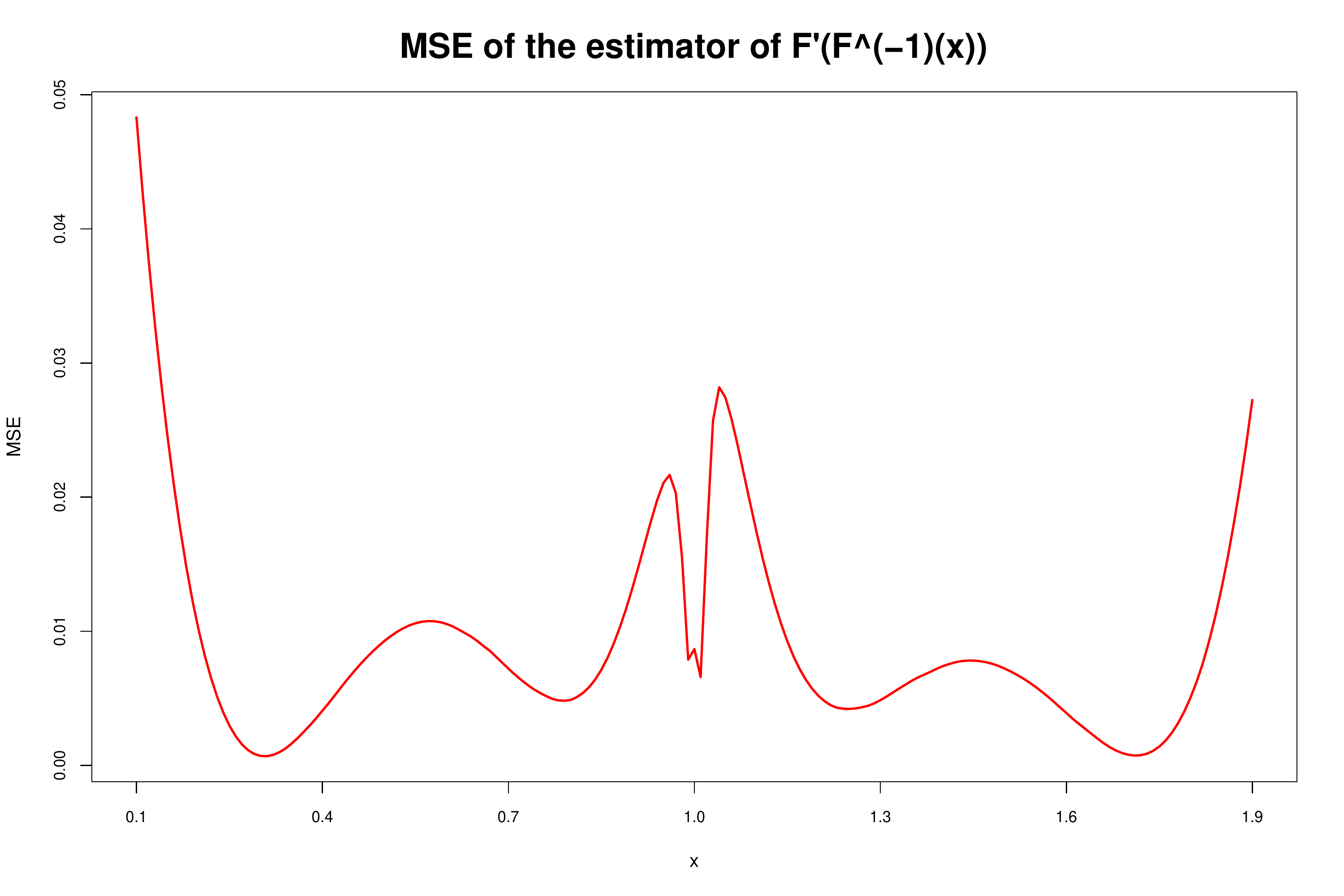} \\
\end{tabular}
}
\caption{Simulated mean squared error of the estimators $\widehat{F}_{mc}$, $\widehat{F}'_{mc}$ and $\widehat{F}_{mc}^{'} \circ \widehat{F}_{mc}^{-1}$ (left to right) computed over 100 simulations runs on the three examples \refeq{eq_simulation_study_f1}-\refeq{eq_simulation_study_f3} (top to bottom).}
\label{fig_simulation_studies_mse}
\end{figure}

We use 100 simulation runs to calculate the pointwise mean squared error (MSE) for the estimators $\widehat{F}_{mc}$ and $\widehat{F}'_{mc}$ evaluated on an equidistant grid of size $2n$, and for the estimator $\widehat{F}_{mc}^{'} \circ \widehat{F}_{mc}^{-1}$ evaluated on an equidistant grid with a step equal to 0.01. Curve of these MSE for the three examples are displayed Fig. \ref{fig_simulation_studies_mse}. Table \ref{tab_simulation_studies_mise} shows the results of the mean integrated squared error (MISE) for the estimators $\widehat{F}_{mc}$, $\widehat{F}'_{mc}$ and $\widehat{F}_{mc}^{'} \circ \widehat{F}_{mc}^{-1}$ in the three examples.

~\\

\begin{table}[!h]
\hspace{-10mm}
\caption{Mean integrated squared error (MISE) of the estimators $\widehat{F}_{mc}$, $\widehat{F}'_{mc}$ and $\widehat{F}_{mc}^{'} \circ \widehat{F}_{mc}^{-1}$, over the 100 simulations for each example.}
\begin{center}
  \begin{tabular}{c|c|c|c}
  \hline
     & $\widehat{F}_{mc}$ & $\widehat{F}'_{mc}$ & $\widehat{F}_{mc}^{'} \circ \widehat{F}_{mc}^{-1}$ \\ \hline
    function $F_{1}$ & 0.00074 & 0.0059 & 0.0033 \\ \hline
    function $F_{2}$ & 0.00084 & 0.0017 & 0.033 \\ \hline
    function $F_{3}$ & 0.00034 & 0.0044 & 0.0092 \\ \hline
  \end{tabular}
\end{center}
\label{tab_simulation_studies_mise}
\end{table}

Fig. \ref{fig_simulation_studies_estimators} and  Fig. \ref{fig_simulation_studies_mse} show the good results of the estimators $\widehat{F}_{mc}$, $\widehat{F}'_{mc}$ and $\widehat{F}_{mc}^{'} \circ \widehat{F}_{mc}^{-1}$, even if we observe some boundary effects at the edge of the interval especially for the derivative estimator $\widehat{F}'_{mc}$. As mentioned Sect. \ref{subsect_properties_of_ssp}, the difficulty in the estimation of the composite function $F' \circ F^{-1}$ is the estimation of the cusp when the function is equal to zero, and that corresponds to a plateau for the function $F$. The comparison between results obtained with the example \refeq{eq_simulation_study_f2} (regression function $F_{2}$ with a small plateau) and the example \refeq{eq_simulation_study_f3} (regression function $F_{3}$ with a large plateau) shows that the cusp is overestimate in the case of a small plateau (in the example \refeq{eq_simulation_study_f2}, the estimator $\widehat{F}_{mc}^{'} \circ \widehat{F}_{mc}^{-1}$ does not cross the axis). This problem results from the mononization step, since the method of Ramsay which is used, provides a strictly increasing estimator $\widehat{F}_{mc}$ that is not appropriate for the estimation of plateaux even if in practice only small plateaux are not very good estimated. However, actually, we have not found a solution for this problem.

\subsection{Application to the real data example}
\label{subsect_application_to_real_data}

\begin{figure}[!h]
\vspace{2 pt}
\centerline{
\begin{tabular}{c}
a. Raw data. \\
\includegraphics[width=10cm]{Figures/speed_vs_distance_raw_data_trajok.pdf} \\
b. Smooth data. \\
\includegraphics[width=10cm]{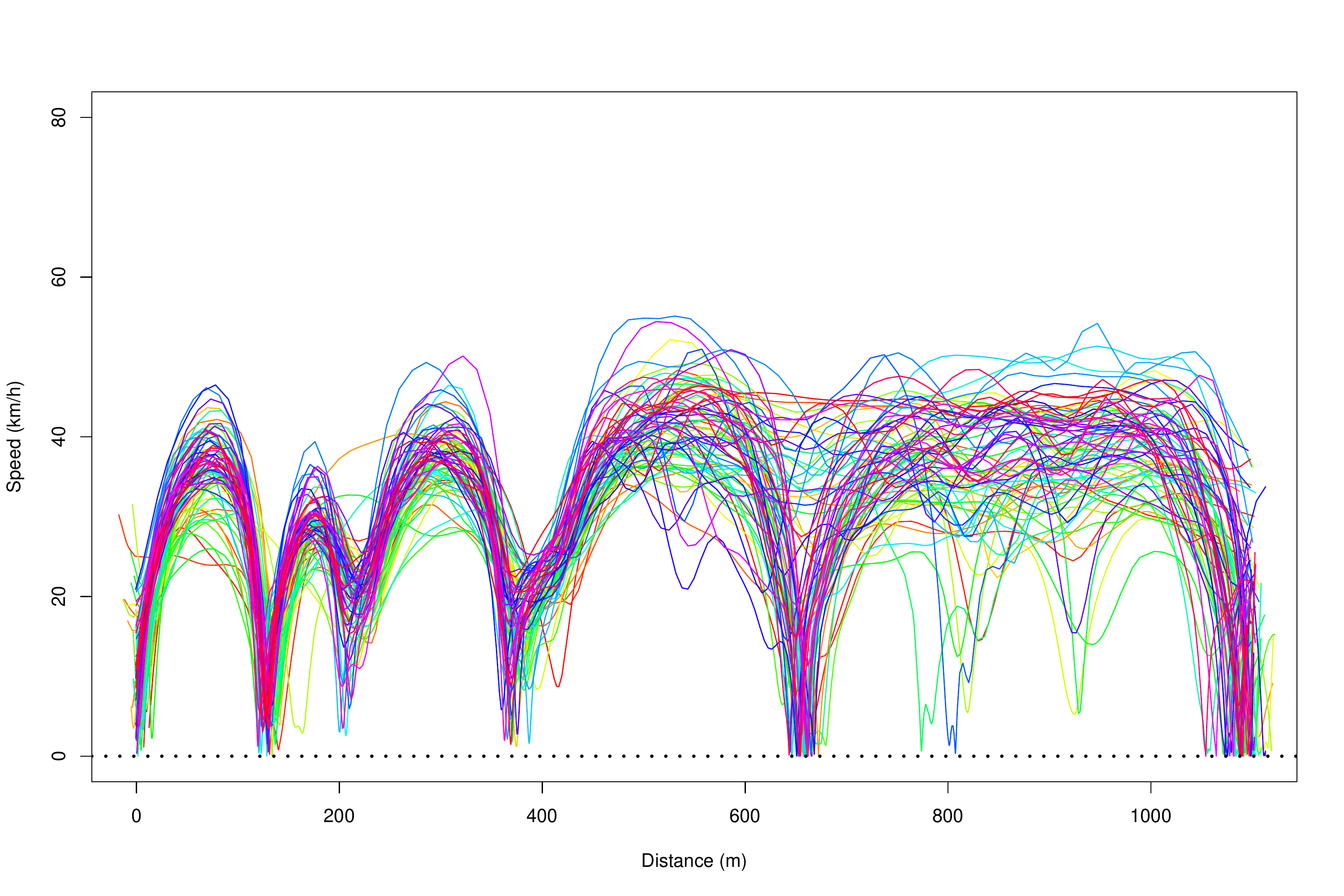} \\
\end{tabular}
}
\caption{Smoothing step on the 78 individual space-speed profiles.}
\label{fig_speed_vs_distance_after_smoothing_and_mono_trajok}
\end{figure}

To illustrate the performance of the smoothing procedure in the estimation of space-speed profiles from noisy position and speed measurements, we applied the method on the real data set presented in Sect. \ref{sect_the_data}. Thus, the smoothing procedure is applied to the data set of 78 individual space-speed profiles illustrated at Fig. \ref{fig_speed_vs_distance_raw_data} and reproduced Fig. \ref{fig_speed_vs_distance_after_smoothing_and_mono_trajok}.a, that are composed of noisy position and speed measurements, as follows:
\begin{enumerate}
  \item A first smoothing step using derivative information, with for each path $j$, $j=1,\ldots,78$, an estimation $\widehat{F}_{\lambda_{j}}(t)$ of each function $F_{j}(t)$ (representing the distance traveled as function of time) with the following parameters:
      \begin{itemize}
        \item an estimation of the variance $\sigma_{x,j}^{2}$ and $\sigma_{v,j}^{2}$ for each path $j$, $j=1,\ldots,78$;
        \item $m=3$ (quintic spline);
        \item an automatic selection of each smoothing parameter $\lambda_{j}$ resulting from the minimization of the GML criterion.
      \end{itemize}
  \item A second smoothing step under monotonicity constraint, corresponding to a monotonization of each estimate $\widehat{F}_{\lambda_{j}}(t)$ obtained at the previous step with the following parameters:
      \begin{itemize}
        \item $m=3$ for the degree of the penalty ;
        \item a selection of each smoothing parameter by trial and errors.
      \end{itemize}
\end{enumerate}
Fig. \ref{fig_speed_vs_distance_after_smoothing_and_mono_trajok}.b illustrates the results of this smoothing procedure and shows the smooth individual space-speed profiles obtained with the transformation $\widehat{F}'_{j,mc}(t)\circ \widehat{F}^{-1}_{j,mc}(t)$. Results are good since some peaks which appear in raw data, and that probably correspond to outliers, are reduced (e.g. blue and orange curves). Missing values are also be corrected by the smoothing procedure. The disadvantage of the step monotonisation mentioned in the simulation study, which causes an over-estimation of speed, appears mainly at the stop sign (short stop) but is less important at the traffic light (long stop). Finally, note that the main difference between Fig. \ref{fig_speed_vs_distance_after_smoothing_and_mono_trajok}.a and Fig. \ref{fig_speed_vs_distance_after_smoothing_and_mono_trajok}.b is that the smoothing procedure allows to reduce the study of these individual speed profiles to a functional
framework since the estimated space-speed profiles belong to the space $\mathcal{E}_{SSP}$ defined in Definition \ref{def_space_speed_profiles}. Some advantages of studying a set of individual space-speed profiles in a functional framework are illustrated in the next sections.

~\\

\section{Curve registration by landmarks alignment}
\label{sect_registration}

The second step of our analysis is to summarize the information contained in a data set of individual space-speed profiles by the construction of the average profile. However, due to phase variation (i.e. horizontal variation) between the individual speed profiles, especially at stops (all vehicles do not stop at the same location), it is necessary to align them in order to obtain an aggregated profile which is representative of the set. \par
The \emph{curve registration} or \emph{curve alignment} problem appears in many areas such as biology, meteorology, pattern recognition... (\citealt{Ramsay1998b}; \citealt{Bigot2006}). Indeed, frequently, observed curves exhibit two types of variability: \emph{amplitude variation} which corresponds to vertical variation, and \emph{phase variation} which corresponds to horizontal variation (\citealt{Ramsay2005}). Then, to build a representative curve of a set of observed curves, it is necessary to correct the phase variation in order to obtain curves with similar features. The curve registration problem consists in finding, for each curve, a warping function and to deform all the curves in order to align them. If the literature about this problem is
relatively large (e.g. \citealt{Kneip1992}; \citealt{Wang1997align}), it is not treated or treated in a simple way in velocity profiles studies.\par
We propose to use the method of landmarks alignment which consists to determine, for each curve, a deformation function so that specific points called landmarks of the registered curves are aligned. Specific points defined as landmarks are generally the positions of maxima, minima, inflection points, or zero crossings. Then, the landmarks registration of $m$ signals $f_{1},\ldots,f_{m}$ defined on the same interval $[0,X]$ can be divided into the five following steps:
\begin{enumerate}
  \item Definition of characteristic points to be used as landmarks (eg, minimum, maximum, zero crossing ...).
  \item Extraction of landmarks $x_{i,1},\ldots,x_{i,K}$ from an observed sequence of each signal $f_{i}$, $i=1\ldots,m$. Note that since observed signals are noisy, the landmarks $x_{i,1},\ldots,x_{i,K}$  are usually extracted from a estimator $\widehat{f}_{i}$ of the signal $f_{i}$.
  \item Identify landmarks reference $x_{0,1},\ldots,x_{0,K}$, i.e. the points at which the curves must match.
  \item Determine deformation functions $h_{1},\ldots,h_{m}$ so that corresponding landmarks are matched, i.e. for all $i=1,\ldots,m$, $h_{i}(x_{0,j})=x_{i,j}$, $j=1,\ldots,K$.
  \item Deformation of the signals using transformations obtained in the previous step. The registered functions $\widetilde{f}_{i}(x)=f_{i}[h^{-1}_{i}(x)]$, $i=1,\ldots,m$, are then aligned at each points $x_{0,1},\ldots,x_{0,K}$.
\end{enumerate}
The deformation functions $h_{i}(x)$, $i=1\ldots,m$, called \emph{warping functions}, must check the following properties:
\begin{itemize}
  \item Initial conditions: $h_{i}(0)=0$, $h_{i}(X)=X$.
  \item Landmarks alignment: $h_{i}(x_{0,j})=x_{i,j}$.
  \item Strict monotonicity: $x_{1}<x_{2}$ implies $h_{i}(x_{1})<h_{i}(x_{2})$ (in order to respect the sequencing of points).
\end{itemize}

\begin{figure}[!h]
\vspace{2 pt}
\centerline{
\begin{tabular}{c}
a. Unregistered space-speed profiles. \\
\includegraphics[width=10cm]{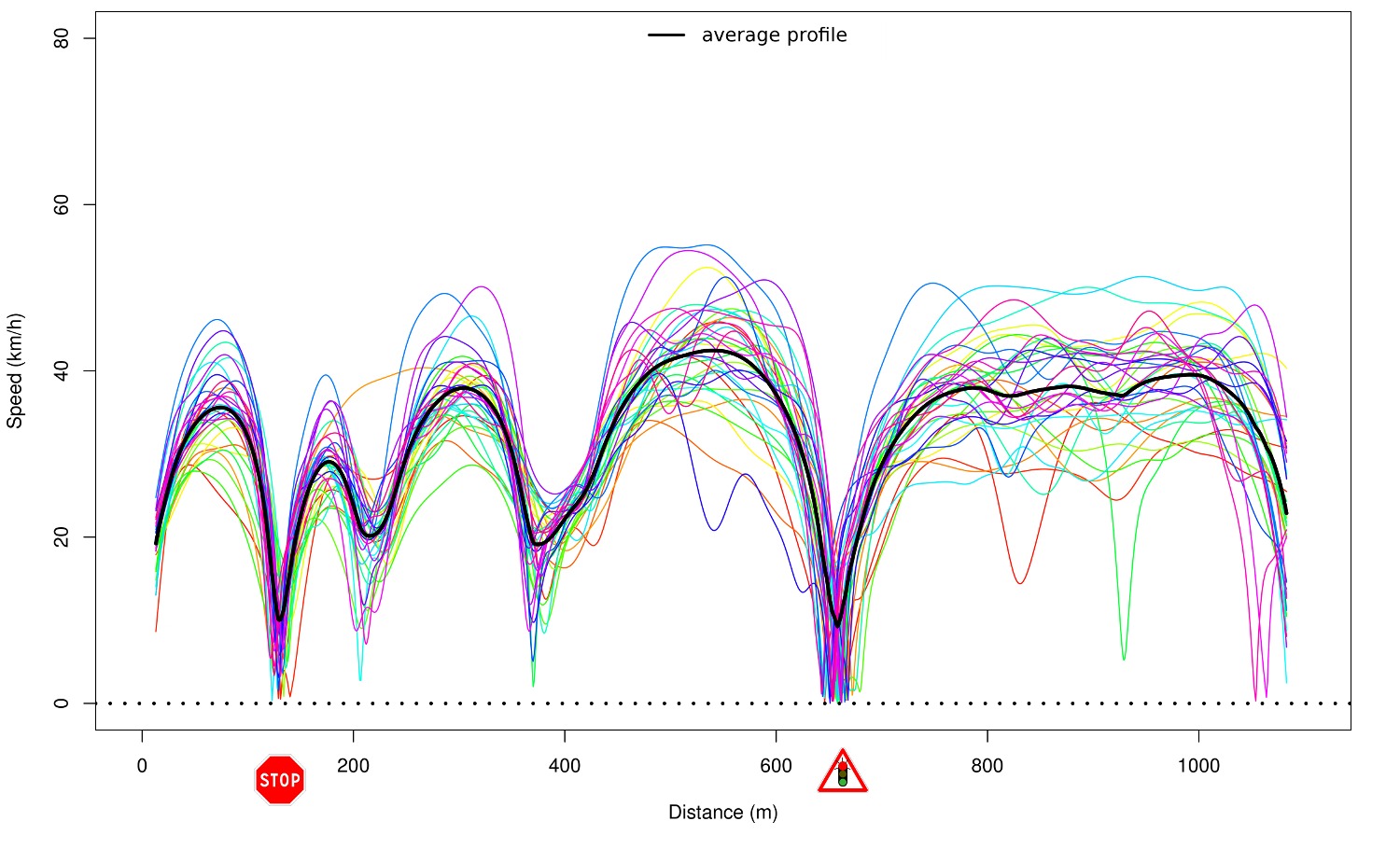} \\
b. Registered space-speed profiles. \\
\includegraphics[width=10cm]{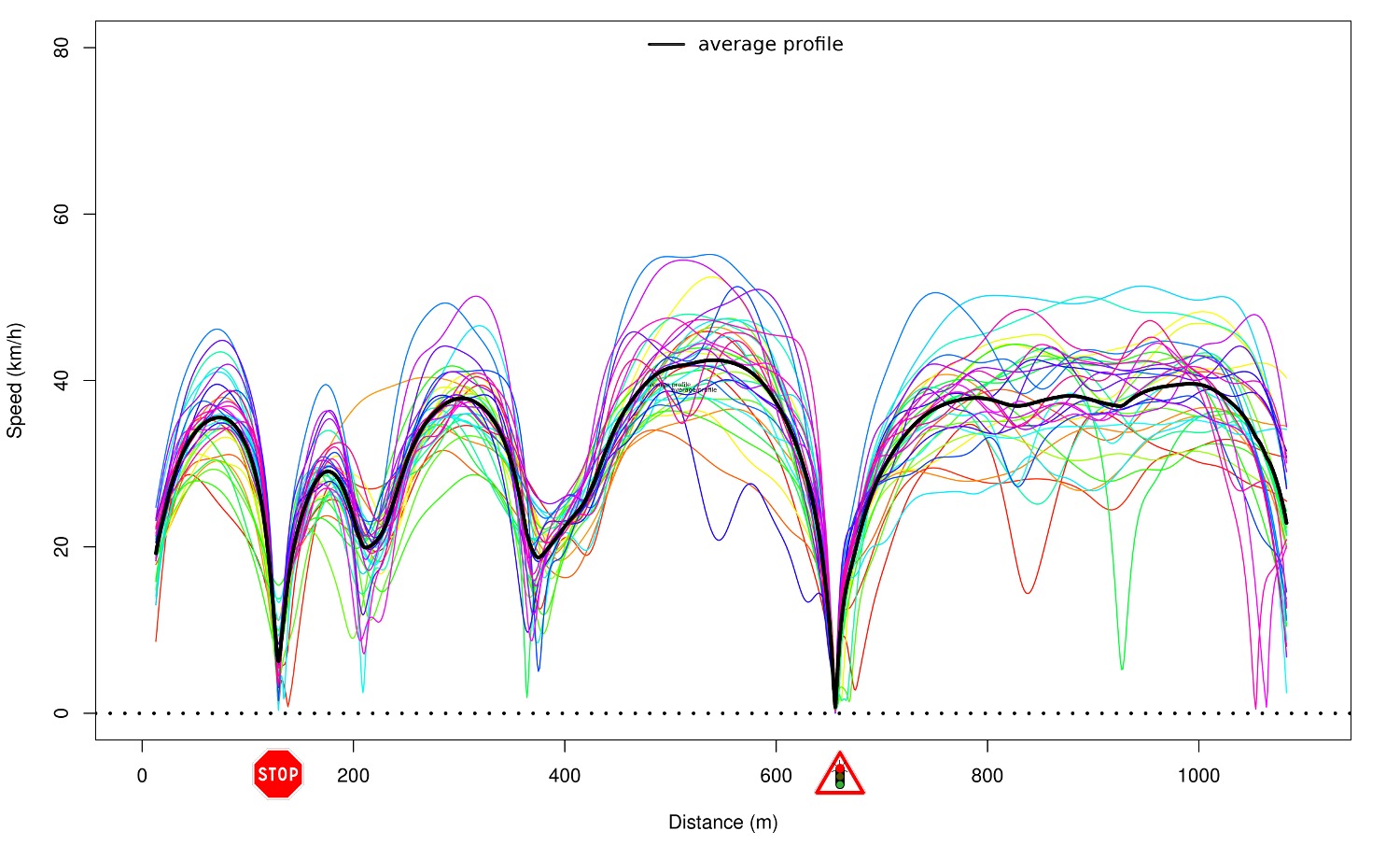} \\
\end{tabular}
}
\caption{Registration of space-speed profiles in the red light case (36 curves). The black curve is the average profile.}
\label{fig_speed_vs_distance_smoothmono_registration_redlight}
\end{figure}
~\\

The method of landmarks alignment is applied to the set of speed profiles illustrated at Fig. \ref{fig_speed_vs_distance_after_smoothing_and_mono_trajok}.b. In order to compare similar speed profiles, we distinguish the two driving situations corresponding to the state of the traffic light (red or green light). Only the red light case will be studied in the following, that represents a sample of 36 individual profiles. We have chosen to define landmarks as the positions of the two elements of the infrastructure that require a stop of the vehicle, namely the stop sign and the red light. Thus, the landmarks, corresponding to zero-crossing (or local minima) at the stop sign and the traffic light positions, are extracted from the estimated space-speed profiles obtained with the smoothing procedure, and are matched with the reference landmarks defined by the average position of vehicle stops at this two elements of the infrastructure. Then, monotone cubic spline interpolation have been determined as warping functions and have been computed with the R function \texttt{splinefun} and the option "monoH.FC". We also impose the condition that the warping functions are linear with a slope equal to one around the stops (we fix an interval of length 100~m around each stop) in order to not too distort the space-speed profiles in the neighborhood of each stop and to obtain "true" space-speed profiles as defined in Definition \ref{def_space_speed_profiles}.\par
Fig. \ref{fig_speed_vs_distance_smoothmono_registration_redlight} compares the unregistered (Fig. \ref{fig_speed_vs_distance_smoothmono_registration_redlight}.a) and the registered (Fig. \ref{fig_speed_vs_distance_smoothmono_registration_redlight}.b) speed profiles in the red light case (36 curves). Fig. \ref{fig_speed_vs_distance_smoothmono_registration_redlight}.a illustrates the fact that averaging unregistered profile results in an average profile (black curve) that is not representative of the set of the individual speed profiles. Indeed, this average profile doesn't equal to zero at the red light unlike all individual profiles. In contrast, Fig. \ref{fig_speed_vs_distance_smoothmono_registration_redlight}.b shows that the average of the registered profiles tends to resemble much more closely most of the individual profiles, and then is a good aggregated speed profile of the sample.

\section{Functional boxplot: a graphical tool to explore the variability of a functional data set}
\label{sect_functional_boxplot}

\begin{figure}[!h]
\vspace{2 pt}
\centerline{
\begin{tabular}{c}
a. Pointwise boxplots. \\
\includegraphics[width=10cm]{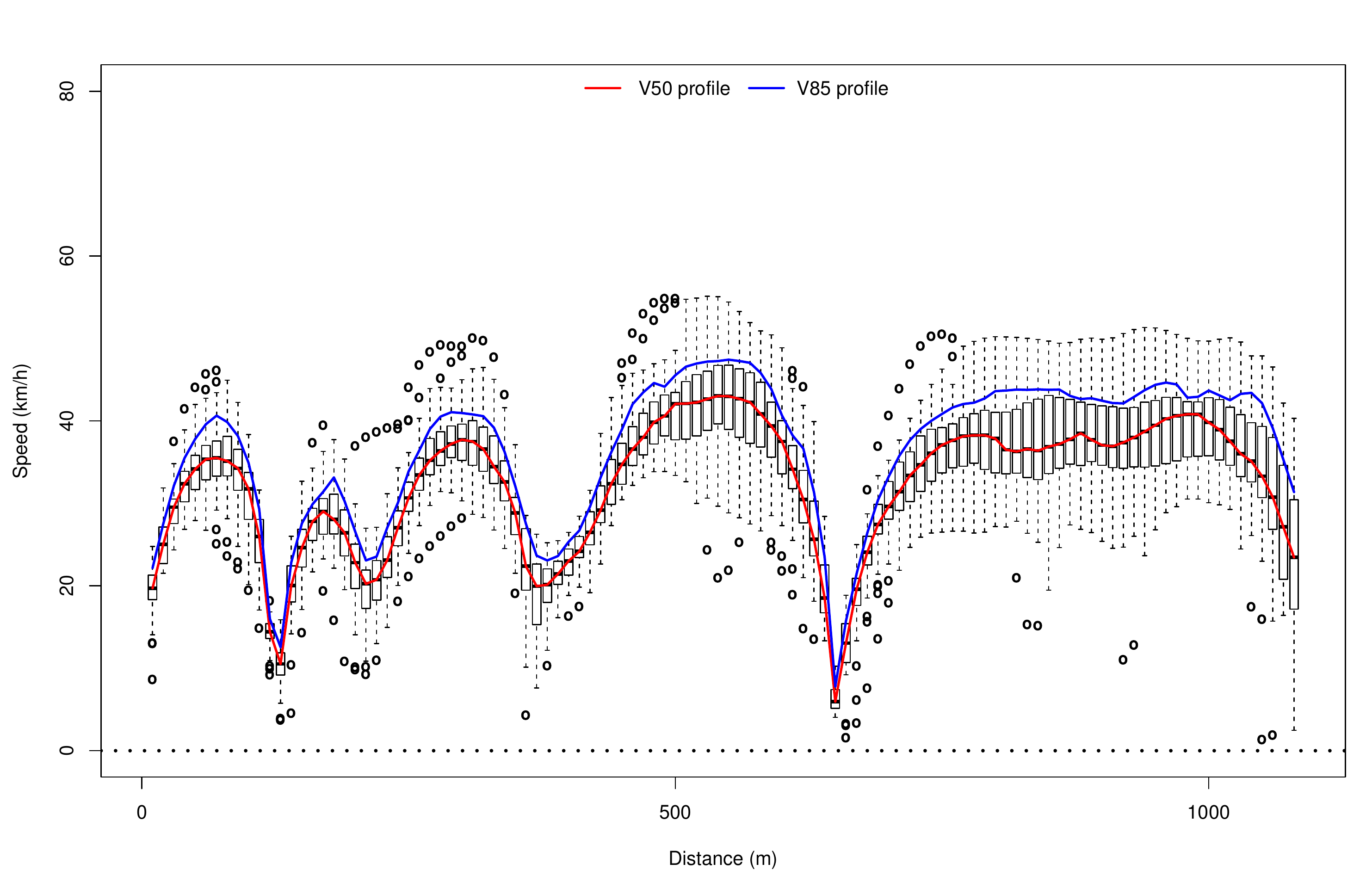} \\
b. Functional boxplots. \\
\includegraphics[width=10cm]{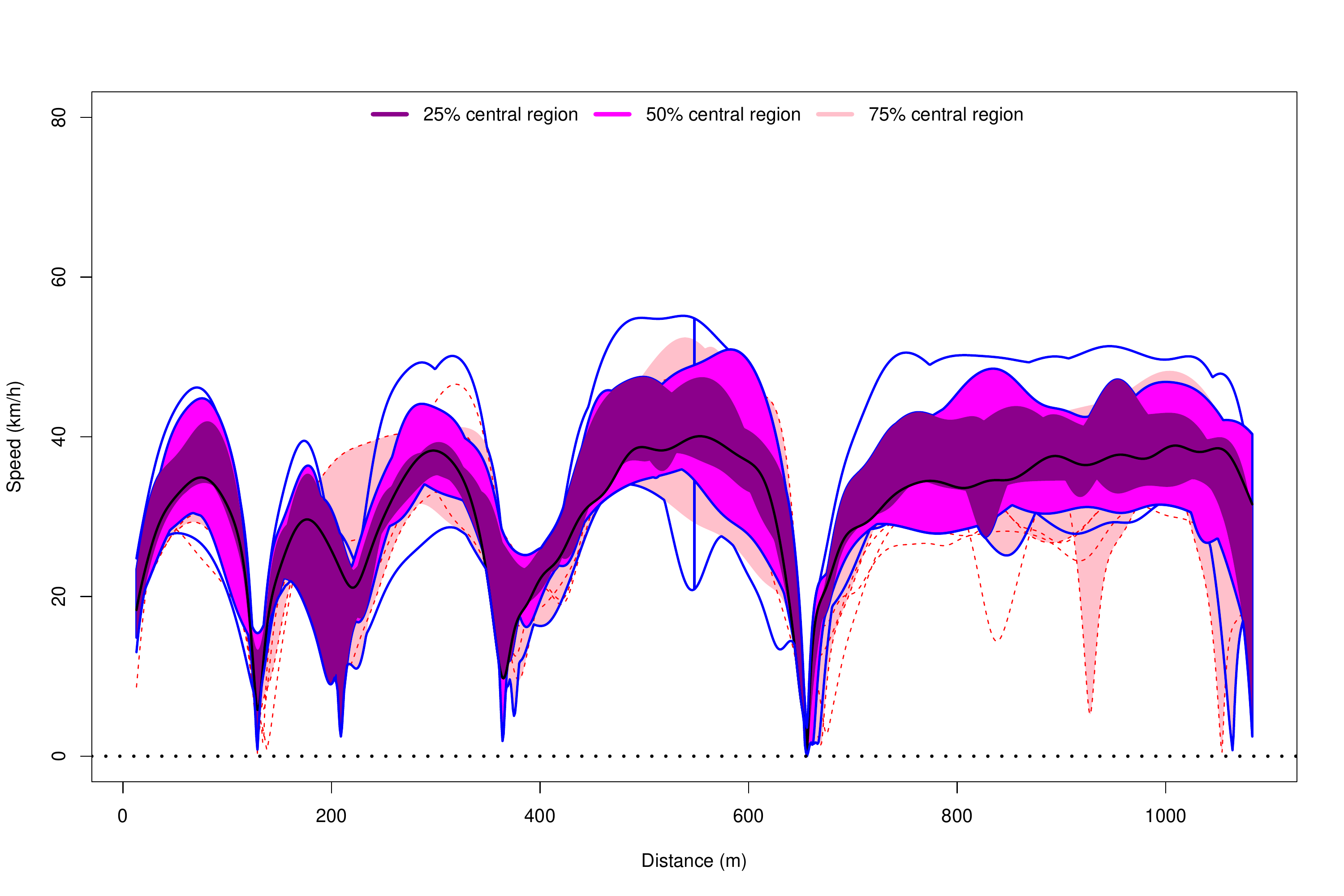} \\
\end{tabular}
}
\caption{Pointwise boxplots ans functional boxplots in the red light case (36 curves).}
\label{fig_speed_vs_distance_pointwise_and_functional_boxplots_redlight}
\end{figure}

Finally, the last step of our analysis is to explore the variability of a set of individual space-speed profiles. Indeed, if the construction of an aggregated speed profile, such as the average profile, leads to a good representation of the actual speeds on a road network section, such an aggregated profile does not reflect the variability between road users. The boxplot proposed by \citet{Tukey1977} is a graphical method used to represent the distribution of univariate data, and can be used to represent speed variations between individuals at a given point. For example, Fig. \ref{fig_speed_vs_distance_pointwise_and_functional_boxplots_redlight}.a represents pointwise boxplots calculated at a regular interval of 10~m in the red light case, with medians connected by a red line (V50 profile) and 85th percentiles connected by a blue line (V85 profile). However, this representation lost the continuous form of the individual profiles, and then the V50 and V85 profiles are not true space-speed profiles as defined in Definition \ref{def_space_speed_profiles} in contrast to the average speed profile obtained at Fig. \ref{fig_speed_vs_distance_smoothmono_registration_redlight}.b. \par
So, we propose to use a graphical tool called functional boxplots, recently developed by \citet{Sun2011}, which extends the notion of boxplots to functional data. This tool is based on the notion of functional depth which generalizes order statistics or ranks to the functional setting. Indeed, the first step to construct a boxplot is the data ordering. But if the notion of order is obvious in the univariate setting, it is much more complicated in the functional setting. This problem has led to the emergence of the concept of functional depth, first introduced for multivariate data (\citealt{Zuo2000}), that provides a measure of "centrality" and "outlyingness" for a function within a sample of curves and allows to order them from center-outward (\citealt{Lopez2009}). The median curve is then the curve with the higher depth. Various examples of functional depth have been proposed in the literature such as the Fraiman and Muniz depth (\citealt{Fraiman2001}), the random projection depth (\citealt{Cuevas2007}) or the band
depth (\citealt{Lopez2009}). If \citet{Sun2011} use the band depth and its modified version for the construction of its functional boxplots, a comparison of the results obtained with various functional depth led us to choose the h-mode depth introduced by \citet{Cuevas2006} and based on the concept of mode. The authors defined a functional mode as the curve most densely surrounded by the rest of curves of the dataset. Thus, the h-modal functional depth of a curve $x_{i}$ with respect
the set of curves $x_{1},\ldots,x_{n}$ is given by:
\begin{equation}
\label{eq_depth_hMD}
MD_{n}(x_{i},h)=\sum_{k=1}^{n}K(\frac{\|x_{i}-x_{k}\|}{h}),
\end{equation}
where $\|.\|$ is an appropriate norm, $K$ is a kernel function, and $h$ is a bandwidth. In practice, the $L^{2}$ norm and the truncated Gaussian kernel are used, and the bandwidth taken is the $15$th percentile of the empirical distribution of $\{\|x_{i}-x_{k}\|,\ i,k=1,\ldots,n\}$. Functional boxplots create with the h-modal depth are illustrated in Fig. \ref{fig_speed_vs_distance_pointwise_and_functional_boxplots_redlight}.b in the red light case. This functional boxplot is composed of the maximum enveloppe (blue curves), the median profile (black curve) which is the most central curve with the highest h-modal depth, the 25\% central region (dark magenta region), the 50\% central region (magenta region) and the 75\% central region (pink region). The red dashed curves are the outlier candidates detected by the 1.5 times the 50\% central region rule (see \citealt{Sun2011}). This functional boxplot have been computed with the function \texttt{fdepth} of the R package \texttt{rainbow} and the function \texttt{fbplot} of the R package \texttt{fda}. \par
The advantage of this graphical tool is that it allows to represent the speed dispersion among individuals on a given road section. These speed corridors allow to distinguish road sections where the speed variability is large and those for which speeds are more homogeneous. Moreover, this tool leads to the extraction of the median profile which depends to the choice of a functional depth, and that can be used as a representative speed profile of the set of the individual speed profiles instead of the average profile.

~\\

\section{Conclusion}
\label{sect_conclusion}

In this paper, we have proposed a functional analysis of a set of space-speed profiles corresponding to speed as function of the distance traveled by the vehicle from an initial point. Thus, a definition of the functional space of these objects was proposed and the study of their mathematical properties has shown the remarkable property of non differentiability at points for which speed is zero and that corresponds to a cusp in the curve. Then, the first step of our analysis was the development of a smoothing procedure in order to be reduced to a functional framework. However, we have shown that the estimation of a space-speed profiles from noisy position and speed measurements was complex and can be reduced to a nonparametric regression problem taking into account two constraints: the use of the derivative information, and a monotonicity constraint. A two-step estimator (smooth, and then monotonize) based on the general theory of thin-plate spline was proposed and computed on simulation studies and on a real data set. If the proposed smoothing procedure presents good results, some limitations appear in the estimation of points for which the function is zero (speed tends to be overestimated at short stops). This point is an important challenge for the future, and the fusion of the two constraints in a single smoothing step will be subject to future research. \par
In a second time, a methodology has been proposed to summarize a set of individual space-speed profiles with an aggregated speed profile. The functional approach allows to use curve registration method in order to correct phase variation, and then to obtain a representative average speed profile with similar features of corresponding individual speed profiles. The method of landmarks alignment, which consists to align specific points of the curves, is applied on a data set where two driving situations corresponding to the state of the traffic light (red or green light) are distinguish. A comparison of the unregistered and the registered speed profiles at stops imposed by the infrastructure in the red light case, as well as the corresponding average profiles, illustrates the interest of the method. The development of unsupervised classification methods to distinguish traffic conditions (free vs congestion) or specific driving conditions (state of traffic lights) could also be subject to future works.\par
Finally, the variability of a set of individual space-speed profiles was explored by the use of functional boxplots, initially proposed by \citet{Sun2011}, which are an extension of the classical boxplots used in the univariate setting. This tool leads to the construction of speed corridors that reflect the variability between road users.


\begin{acknowledgements}

The authors would like to gratefully thank Christine Thomas-Agnan from the University of Toulouse I for discussions about RKHS and the general smoothing problem, and for her help in the use of the derivative information in the smoothing procedure presented in this article. We thank Yuedong Wang from the University of California for his help in the implementation of the smoothing procedure with the R package \texttt{assist}, and Jeremie Bigot from the Toulouse Mathematics Institute for discussions about the present study. This work was funded by the French Institute of Science and Technology for Transport, Development and Networks (IFSTTAR).

\end{acknowledgements}



\appendix


\section{Proofs of properties given in Section \ref{subsect_properties_of_ssp}}
\label{appendix_proofs_properties_of_ssp}

\begin{pfof}{Theorem \ref{th_continuity_ssp}}
~\\
Let $x_{0} \in [0,x_{f}]$. There are two distinct cases :\\
$1^{st}$ case : $x_{0}$ is a point of continuity of $F^{-1}$. Then by composition of two continuous functions, we deduce that $F' \circ F^{-1}$ is continuous at $x_{0}$.\\
$2^{nd}$ case : $x_{0}$ is a point of discontinuity of $F^{-1}$. We begin by demonstrating the following lemma:

\begin{lemma}
\label{lemme_cas_vitesse_nulle}
Let $x_{0} \in [0,x_{f}]$  a point of discontinuity of $F^{-1}$. Then the speed is zero at this point, i.e. $v_{S}(x_{0})=F' \circ F^{-1} (x_{0})=0$.
\end{lemma}

~\\

This lemma can be proved easily. Indeed, if $x_{0}$ is a point of discontinuity of $F^{-1}$, then there is a close interval $[t^{-},t^{+}]$ where $F$ is constant and equal to $x_{0}$, and by definition of $F^{-1}$, $F^{-1}(x_{0})=t^{-}$. This implies that $F'_{+}(t^{-})=0$ where $F'_{+}(t^{-})$ is the right derivative of $F$ at $t^{-}$, and as it was assumed that $F$ was differentiable, we also have $F'_{-}(t^{-})=0$ where $F'_{-}(t^{-})$ is the left derivative of $F$ at $t^{-}$. Finally, $F'(t^{-})=0$, and therefore $F' \circ F^{-1} (x_{0})=0$ which ends the proof of the lemma \ref{lemme_cas_vitesse_nulle}. \\
Now, we study the one-sided limit of $v_{S}=F' \circ F^{-1}$ at $x_{0}$. $(v_{S})_{-}(x_{0})=\lim\limits_{\substack{x \to x_{0} \\ x<x_{0}}} v_{S}(x)=\lim\limits_{\substack{x \to x_{0} \\ x<x_{0}}} F'~\circ~F^{-1}(x)$. When $x \to x_{0}$ by lower values, $t \to t^{-}$ by lower values, and $\lim\limits_{\substack{t \to t^{-} \\ t<t^{-}}}F'(t)=0$ since $F'(t)=0$ on $[t^{-},t^{+}]$ and $F'$ is continuous at $t^{-}$. So, we deduce that $(v_{S})_{-}(x_{0})=0$. Similarly, $(v_{S})_{+}(x_{0})=\lim\limits_{\substack{x \to x_{0} \\ x>x_{0}}} v_{S}(x)=\lim\limits_{\substack{x \to x_{0} \\ x>x_{0}}} F' \circ F^{-1}(x)$. When $x \to x_{0}$ by upper values, $t \to t^{-}$ by upper values, and $\lim\limits_{\substack{t \to t^{-} \\ t>t^{-}}}F'(t)=0$ since $F'(t)=0$ on $[t^{-},t^{+}]$. So, we deduce that $(v_{S})_{+}(x_{0})=0$. Hence, using Lemma \ref{lemme_cas_vitesse_nulle}, we conclude that $v_{S}$ is continuous at $x_{0}$.
\end{pfof}

~\\
\begin{pfof}{Theorem \ref{th_differentiability_ssp}}
~\\
$1^{st}$ case: Assume that $F$ satisfies the assumptions $(H_{1})$.\\
Let $x_{0}$ such that $t_{0}=F^{-1}(x_{0})$. Since $F'(t_{0})=0$, then $v_{S}(x_{0})=0$, i.e. $x_{0}\in H_{0}$. Under the assumptions $(H_{1})$, we can apply the Taylor-Young's formula to $F'$: For all $\theta$ in a neighborhood of $t_{0}$, $F'(t_{0}+\theta)=F'(t_{0})+\theta F''(t_{0})+\frac{\theta^{2}}{2}F'''(t_{0})+\theta^{2}\varepsilon(\theta)$, where $\varepsilon(\theta)\rightarrow 0$ when $\theta~\rightarrow~0$. But since $F'(t_{0})=0$, if we had $F''(t_{0}) \neq 0$, then $F'$ would change sign at $t_{0}$, which contradicts the strict monotonicity of $F$. Therefore $F''(t_{0})=0$. So, $F'(t_{0}~+~\theta)~\underset{\theta \rightarrow 0}{\sim}~\frac{\theta^{2}}{2}F'''(t_{0})$ (since it is assumed that $F'''(t_{0}) \neq 0$). \\ Let $h~=~F(t_{0}~+~\theta)~-~F(t_{0})$. We apply the Taylor-Young's formula to $F$ :\\
$h=F(t_{0}+\theta)-F(t_{0})=\theta F'(t_{0})+\frac{\theta^{2}}{2}F''(t_{0})+ \frac{\theta^{3}}{6}F'''(t_{0}) + \theta^{3}\varepsilon^{'}(\theta)$ where $\varepsilon^{'}(\theta)\rightarrow 0$ when $\theta \rightarrow 0$. In order to study the differentiability of $v_{s}$ at $x_{0}$, we define the following growth rates:\\
$\frac{v_{S}(x_{0}+h)-v_{S}(x_{0})}{h}=\frac{F'(t_{0}+\theta)-F'(t_{0})}{F(t_{0}+\theta)-F(t_{0})} \underset{\theta \rightarrow 0}{\sim} \frac{\frac{\theta^{2}}{2}F'''(t_{0})}{\frac{\theta^{3}}{6}F'''(t_{0})} = \frac{3}{\theta}$. This growth rate has no limit when $\theta \rightarrow 0$, but this does not prove that it has also no limit when $h \rightarrow 0$. \\
We will prove this by contradiction. Assume that $\frac{v_{S}(x_{0}+h)-v_{S}(x_{0})}{h} \underset{h \rightarrow 0}{\rightarrow} \ell \in \mathds{R}$. Then, by definition, \\
$\forall \varepsilon>0$, $\exists \alpha>0$ such that $|h|<\alpha \Rightarrow |\frac{v_{S}(x_{0}+h)-v_{S}(x_{0})}{h}-\ell |<\varepsilon$.\\
But since $F$ is continuous at $t_{0}$, $\exists \beta >0$ such that $|t-t_{0}|<\beta \Rightarrow |F(t)-F(t_{0})|<\alpha$, or similarly $|\theta| < \beta \Rightarrow |\underbrace{F(t_{0}+\theta)-F(t_{0})}_{h}|<\alpha$.\\
Hence, $\forall \varepsilon>0$, $\exists \beta>0$ such that $|\theta|<\beta \Rightarrow |\frac{v_{S}(x_{0}+h)-v_{S}(x_{0})}{h}-\ell |<\varepsilon$. This means that the growth rate has a limit $\ell \in \mathds{R}$ when $\theta \rightarrow 0$, which is a contradiction. Hence, under the assumptions $(H_{1})$, $v_{S}$ is not differentiable at $x_{0}$.\\
\\

$2^{nd}$ case: Assume that $F$ satisfies the assumptions $(H_{2})$.\\
As in the first case, we define $x_{0}$ such that $t_{0}=F^{-1}(x_{0})$. The graph of $G$ :
\begin{itemize}
  \item coincides with $F$ on $[0,t_{0}]$,
  \item is deduced from the graph of $F$ by the translation vector $(t_{0}-t_{1})\overrightarrow{i}$ on $[t_{0},T-(t_{1}-t_{0})]$.
\end{itemize}
Thus, the graph of $G$ is similar to the graph of $F$ but removing the time period $[t_{0},t_{1}]$ for which the function is constant. So, the same growth rate occurs at $x_{0}$, and if $v_{S}$ is not differentiable at $x_{0}$ for one, it is not for the other. In other words, the results of the first case where $F'=0$ at one point $t_{0}$ extend to the more general case where $F'$ is zero on an interval $[t_{0},t_{1}]$ ($t_{0} \neq t_{1}$), subject to the assumptions $(H_{1})$ on $G$.
\end{pfof}


\bibliographystyle{spbasic}      
\bibliography{biblioTheseArtCompStat}   

\end{document}